\documentclass[12pt,preprint]{aastex}
\usepackage[intlimits]{amsmath}

\newcommand{\R}{{\mathbb R}}
\newcommand{\C}{{\mathbb C}}

\begin{document}

\title{New Orbits for the $n$-Body Problem}

\author{Robert J. Vanderbei}
\affil{Operations Research and Financial Engineering, 
Princeton University}
\email{rvdb@princeton.edu}

\begin{abstract}
In this paper, we consider minimizing the action functional as a
method for numerically discovering periodic solutions to the $n$-body
problem.  With this method, we can find a large number of choreographies
and other more general solutions.  We show that most of the solutions
found, including all but one of the choreographies, are unstable.
It appears to be much easier to find unstable solutions to the $n$-body
problem than stable ones.  Simpler solutions are more likely to be stable
than exotic ones.
\end{abstract}


\section{Least Action Principle}

Given $n$ bodies, let
$m_j$ denote the mass and 
$z_j(t)$ denote the position in $\R^2 = \C$ of body $j$ at time $t$.
The {\em action functional} is a mapping from the space of all
trajectories, $z_1(t),z_2(t),\ldots,z_n(t)$, $0 \le t \le 2 \pi$, 
into the reals.
It is defined as the integral over one period of the
kinetic minus the potential energy:
\[
    A = \int_0^{2\pi} \left( 
    			\sum_j \frac{m_j}{2} \| \dot{z}_j \|^2 
			+
			\sum_{j,k:k<j} \frac{m_j m_k}{\|z_j - z_k\|}
                 \right) dt.
\]

Stationary points of the action function are trajectories that satisfy the
equations of motions, i.e., Newton's law gravity.  To see this, we compute
the first variation of the action functional,
\begin{eqnarray*}
    \delta A 
    & = & 
    \int_0^{2\pi}
    \sum_{\alpha} 
    \left(
        \sum_j m_j \dot{z}_j^{\alpha} \dot{\delta z}_j^{\alpha}
	-
	\sum_{j,k:k<j} 
	m_j m_k
	\frac{
	  (z_j^{\alpha}-z_k^{\alpha})(\delta z_j^{\alpha}-\delta z_k^{\alpha})
	}{
	  \| z_j - z_k \|^3
	}
    \right)
    dt
    \\
    & = &
    -
    \int_0^{2\pi} \sum_j \sum_{\alpha} 
    \left(
	m_j
	\ddot{z}_j^{\alpha} 
	+
	\sum_{k:k \ne j} 
	m_j m_k
	\frac{
	z_j^{\alpha}-z_k^{\alpha}
	}{
	\| z_j - z_k \|^3
	}
    \right)
    \delta z_j^{\alpha}
    dt, 
\end{eqnarray*}
and set it to zero.  We get that
\begin{equation}
    m_j
    \ddot{z}_j^{\alpha} 
    = 
    - \sum_{k: k \ne j} m_j m_k 
                        \frac{z_j^{\alpha}-z_k^{\alpha}}{\| z_j-z_k \|^3},
    \qquad
    j=1,2,\ldots,n, \quad \alpha=1,2
    \label{52}
\end{equation}

Note that if $m_j = 0$ for some $j$, then the first order optimality condition
reduces to $0 = 0$, which is {\em not} the equation of motion for a massless
body.  Hence, we must assume that all bodies have strictly positive mass.

\section{Periodic Solutions}

Our goal is to use numerical optimization to minimize the action functional
and thereby find periodic solutions to the $n$-body problem.  Since we are
interested only in periodic solutions, we express all trajectories in terms of
their Fourier series:
\[
    z_j(t) = \sum_{k=-\infty}^{\infty} \gamma_k e^{ikt}, \qquad
    \gamma_k \in \C .
\]
Abandoning the efficiency of complex-variable notation, we can 
write the trajectories with components
$
    z_j(t) = (x_j(t),y_j(t)) 
$
and
$
    \gamma_k = (\alpha_k, \beta_k).
$
So doing, we get
\begin{eqnarray*}
    x(t) & = & a_0 + \sum_{k=1}^{\infty} 
               \left( a_k^c \cos(kt) + a_k^s \sin(kt) \right) \\
    y(t) & = & b_0 + \sum_{k=1}^{\infty} 
               \left( b_k^c \cos(kt) + b_k^s \sin(kt) \right) 
\end{eqnarray*}
where
\begin{eqnarray*}
    a_0 = \alpha_0, \qquad 
    a_k^c & = & \alpha_{k}+\alpha_{-k}, \qquad
    a_k^s = \beta_{-k}-\beta_{k},  
    \\
    b_0 = \beta_0, \qquad 
    b_k^c & = & \beta_{k}+\beta_{-k}, \qquad
    b_k^s = \alpha_{k}-\alpha_{-k}.  
\end{eqnarray*}
Since we plan to optimize over the space of trajectories,
the parameters $a_0$, $a_k^c$, $a_k^s$, $b_0$, $b_k^c$, and $b_k^s$ are 
the decision variables in our optimization model.
The objective is to minimize the action functional.  

{\sc ampl} is a small programming language designed for the efficient
expression of optimization problems \cite{FGK93}.  Figure \ref{fig:50} shows the {\sc
ampl} program for minimizing the action functional.  

Note that the action
functional is a nonconvex nonlinear functional.  Hence, it is expected to
have many local extrema and saddle points.  We use the author's local
optimization software called {\sc loqo} (see \cite{SOR9708}, \cite{Van97d})
to find local minima in a neighborhood
of an arbitrary given starting trajectory.  
One can provide either specific initial
trajectories or one can give random initial trajectories.  The four lines
just before the call to {\tt solve} in Figure \ref{fig:50} show how to specify
a random initial trajectory.
Of course, {\sc ampl} provides capabilities of printing answers in any format
either on the standard output device or to a file.  For the sake of brevity
and clarity, the print statements are not shown in Figure \ref{fig:50}.  {\sc
ampl} also provides the capability to loop over sections of code.  This is
also not shown but the program we used has a loop around the four
initialization statements, the call to solve the problem, and the associated
print statements.  In this way, the program can be run once to solve for 
a large number of periodic solutions.

\subsection{Choreographies}
Recently, \citet{CM00} introduced a new family of solutions to the $n$-body
problem called choreographies.  A {\em choreography} is defined as a solution
to the $n$-body problem in which all of the bodies share a common orbit and
are uniformly spread out around this orbit.  Such trajectories are even easier
to find using the action principle.  Rather than having a Fourier series for
each orbit, it is only necessary to have one master Fourier series and to
write the action functional in terms of it.  Figure \ref{fig:51} shows the {\sc
ampl} model for finding choreographies.

\section{Stable vs. Unstable Solutions}

Figure \ref{fig:fig1} shows some simple choreographies found by minimizing the
action functional using the {\sc ampl} model in Figure \ref{fig:51}.  
The famous $3$-body figure eight, first discoverd by \citet{Mor93}
and later analyzed by \citet{CM00}, 
is the first one shown---labeled FigureEight3.
It is easy to find choreographies of arbitrary complexity.  In fact, it is
not hard to rediscover most of the choreographies given in \cite{CGMS01}, 
and more,
simply by putting a loop in the {\sc ampl} model and finding various
local minima by using different starting points.

However, as we discuss in a later section, simulation makes it
apparent that, with the sole exception of FigureEight3, all
of the choreographies we found are unstable.  And, the more intricate
the choreography, the more unstable it is.  Since the only
choreographies that have a chance to occur in the real world are stable
ones, many cpu hours were devoted to searching
for other stable choreographies.  So far, none have been found.
The choreographies shown in Figure \ref{fig:fig1}
represent the ones closest to being stable.  

Given the difficulty of finding stable choreographies, it seems interesting
to search for stable nonchoreographic solutions using, for example,
the {\sc ampl} model from Figure \ref{fig:50}.
The most interesting 
such solutions are shown in Figure \ref{fig:fig2}.  The one
labeled Ducati3 is stable as are Hill3\_15 and the three DoubleDouble solutions.
However, the more exotic solutions (OrthQuasiEllipse4, Rosette4, PlateSaucer4,
and BorderCollie4) are all unstable.

For the interested reader, a {\sc java} applet can be found at 
\cite{GravityApplet}
that allows one to watch the dynamics of each of the systems presented
in this paper (and others).  This applet actually integrates
the equations of motion.  If the orbit is unstable it becomes very obvious
as the bodies deviate from their predicted paths.

\subsection{Ducati3 and its Relatives}

The Ducati3 orbit
first appeared in \cite{Mor93} and has been independently rediscovered
by this author, Broucke \cite{Bro03}, and perhaps others.  
Simulation reveals it to be a stable
system.  The {\sc java} applet at \cite{GravityApplet} allows one to rotate
the reference frame as desired.  By setting the rotation to counter
the outer body in Ducati3, one discovers that the other two bodies
are orbiting each other in nearly circular orbits.  In other words,
the first body in Ducati3 is executing approximately a circular orbit,
$z_1(t) = -e^{it}$, the second body is oscillating back and forth roughly
along the
$x$-axis, $z_2(t) = \cos(t)$, and the third body is oscillating up and down
the $y$-axis, $z_3(t) = i \sin(t)$.  Rotating so as to fix the first body
means multiplying by $e^{-it}$:
\begin{eqnarray*}
    \bar{z}_1(t) & = & e^{-it} (-e^{it}) = -1 \\
    \bar{z}_2(t) & = & e^{-it} \cos(t) = (1+ e^{-2it})/2 \\
    \bar{z}_2(t) & = & e^{-it} i \sin(t) = (1- e^{-2it})/2 .
\end{eqnarray*}
Now it is clear that bodies 2 and 3 are orbiting each other at half the
distance of body 1.  So, this system can be described as a Sun, Earth, Moon
system in which all three bodies have equal mass and in which one (sidereal)
month equals one year.  The synodic month is shorter---half a year.

This analysis of Ducati3 suggests looking for other stable solutions of the
same type but with different resonances between the length of a month and a
year.  Hill3\_15 is one of many such examples we found.  In Hill3\_15, there are
15 sidereal months per year.  Let Hill3\_$n$ denote the system in which
there are $n$ months in a year.  All of these orbits are easy to calculate
and they all appear to be stable.  This success suggests going in the other
direction.  Let Hill3\_$\frac{1}{n}$ denote the system in which there are $n$
years per month.  We computed Hill3\_$\frac{1}{2}$ and found it to be unstable.
It is shown in Figure \ref{fig:fig4}.

In the preceding discussion, we decomposed these Hill-type systems into two
$2$-body problems:  the Earth and Moon orbit each other while their center of 
mass orbits the Sun.  This suggests that we can find
stable orbits for the $4$-body problem by splitting the Sun into a binary star.
This works.  The orbits labeled DoubleDouble$n$ are of this type.  As already
mentioned, these orbits are stable.

Given the existence and stability of FigureEight3, one often is asked
if there is
any chance to observe such a system among the stars.  The answer is that it
is very unlikely since its existence depends crucially on the masses being
equal.  The Ducati and Hill type orbits, however, are not constrained to 
have their masses be equal.  Figure \ref{fig:fig3} shows several Ducati-type
orbits in which the masses are not all equal.  All of these orbits are stable.
This suggests that stability is common for Ducati and Hill type orbits.
Perhaps such orbits can be observed.

\section{Limitations of the Model}

The are certain limitations to the approach articulated above.
First, the Fourier series is an infinite sum that
gets truncated to a finite sum in the computer model.  Hence, the trajectory
space from which solutions are found is finite dimensional.

Second, the integration is replaced with a Riemann sum.  If the discretization
is too coarse, the solution found might not correspond to a real solution to
the $n$-body problem.  The only way to be sure is to run a simulator.

Third, as mentioned before, all masses must be positive.  If there is a zero
mass, then the stationary points for the action function, which satisfy
\eqref{52}, don't necessarily satisfy the equations of motion given by Newton's
law.

Lastly, the model, as given in Figure \ref{fig:50},
can't solve 2-body problems with eccentricity.
We address this issue in the next section.

\section{Elliptic Solutions}

An ellipse with semimajor axis $a$, semiminor axis $b$, and having its
left focus at the origin of the coordinate system is given parametrically
by:
\[
    x(t) = f + a \cos t , \qquad y(t) = b \sin t ,
\]
where $f = \sqrt{a^2 - b^2}$ is the distance from the focus to the center of
the ellipse.

However, this is {\em not} the trajectory of a mass in the $2$-body
problem.  Such a mass will travel faster around one focus than around the
other.  To accomodate this, 
we need to introduce a time-change function $\theta(t)$:
\[
    x(t) = f + a \cos \theta(t) , \qquad y(t) = b \sin \theta(t) .
\]
This function $\theta$ must be increasing and must satisfy $\theta(0) = 0$
and $\theta(2\pi) = 2 \pi$.

The optimization model can be used to find (a discretization of) $\theta(t)$
automatically by changing {\tt param theta} to {\tt var theta} and adding
appropriate monotonicity and boundary constraints.  In this manner, more
realistic orbits can be found that could be useful in real space missions.

In particular, 
using an eccentricity $e = f/a = 0.0167$ and appropriate Sun and Earth masses,
we can find a periodic 
Hill-Type satellite trajectory in which the satellite orbits the
Earth once per year.


\section{Sensitivity Analysis}
The determination of stability vs. instability mentioned the previous
sections was done empirically by simulating the orbits with a 
integrator and very small step sizes.  Two integrators were used: a midpoint
integrator and a $4$-th order Runge-Kutta integrator.  Orbits that are claimed
to be stable were run for several hours of cpu time (which corresponds to
many thousands of orbits) without falling apart.  Orbits that are claimed
to be unstable generally became obviously so in just a few seconds of cpu
time, which corresponds to only a few full orbits.  In this section, we
describe a Floquet analysis of stability and present this measure of stability
for the various orbits found.

For simplicity, in this section we assume that all masses are equal to one.
Let 
$
    \xi^*(t) = ( z^*(t) , \dot{z}^*(t) )
$
be a particular solution to
\[
    \dot{\xi} = A(\xi)
\]
where
\[
    A\left( z(t) , \dot{z}(t) \right) 
    =
    ( \dot{z}(t) , a(z(t)))
\]
and
\[
    a(z) 
    = 
    ( a_1(z) , \ldots , a_n(z) )
\]
and
\[
    a_j(z) = - \sum_{k:k \ne j} \frac{z_j-z_k}{\| z_j-z_k \|^2 },
    \quad j=1,2,\ldots,n .
\]
Consider a nearby solution $\xi(t)$:
\begin{eqnarray*}
    \dot{\xi}(t) 
    & = & A(\xi(t)) \\
    & \approx & A(\xi^*(t)) + A'(\xi^*(t))(\xi(t)-\xi^*(t))
    \\
    & = &
    \dot{\xi}^*(t) + A'(\xi^*(t))(\xi(t)-\xi^*(t)) .
\end{eqnarray*}
Put $\Delta \xi = \xi - \xi^*$. Then
$
    \dot{\Delta \xi} = A'(\xi^*(t)) \Delta \xi .
$
A finite difference approximation yields
\begin{eqnarray*}
    \Delta \xi(t+h) & = & \Delta \xi(t) + h A'(\xi^*(t)) \Delta \xi(t) \\
                    & = & \left( I + h A'(\xi^*(t)) \right) \Delta \xi(t) .
\end{eqnarray*}
Iterating around one period, we get:
\[
    \Delta \xi(T) 
    = 
    \left(
    \prod_{i=0}^{n-1} \left( I + h A'(\xi^*(t_i)) \right)
    \right)
    \Delta \xi(0) ,
\]
where $h = T/n$ and $t_i = iT/n$.

The following perturbations, which are associated with invariants of the 
physical laws, are unimportant in calculating $\Delta \xi(T)$:

\[
  \left[ \begin{array}{c} \Delta z \\ \Delta \dot{z} \end{array} \right] 
  = 
  \left[ \begin{array}{c} e_1 \\ e_1 \\ e_1 \\ 0 \\ 0 \\ 0 \end{array} \right],
  \quad 
  \left[ \begin{array}{c} e_2 \\ e_2 \\ e_2 \\ 0 \\ 0 \\ 0 \end{array} \right],
  \quad 
  \left[ \begin{array}{c} 0 \\ 0 \\ 0 \\ e_1 \\ e_1 \\ e_1 \end{array} \right],
  \quad 
  \left[ \begin{array}{c} 0 \\ 0 \\ 0 \\ e_2 \\ e_2 \\ e_2 \end{array} \right],
  \quad 
  \left[ \begin{array}{c} 
      R z_1 \\ R z_2 \\ R z_3 \\ R \dot{z}_1 \\ R \dot{z}_2 \\ R \dot{z}_3 
  \end{array} \right],
  \quad 
  \frac{1}{2}
  \left[ \begin{array}{c} 
      -3 \dot{z}_1 + 2 z_1 \\ -3 \dot{z}_1 + 2 z_1 \\ -3 \dot{z}_1 + 2 z_1 \\ 
      -3 a_1 -   \dot{z}_1 \\ -3 a_2 -   \dot{z}_2 \\ -3 a_3 -   \dot{z}_3 
  \end{array} \right],
\]
where $R$ denotes rotation by $90^{\circ}$.
The first two of these perturbations correspond to {\em translation}.
The next two correspond to {\em moving frame of reference}
and the last two correspond to {\em rotation}, and {\em dilation}.  
Dilation is explained below. 
Of course, all positions and velocities are evaluated at $t=0$.  
Vector $e_i$ denotes the $i$-th unit vector in $\R^2$.

Consider spatial dilation by $\rho$ together with a temporal dilation by
$\theta$:
\[
  Z_j(t) = \rho z_j (t/\theta) .
\]
Given that the $z_j$'s are a solution, it is easy to check that
\[
    \ddot{Z}_j(t) 
    =
    - \frac{\rho^3}{\theta^2} 
    \sum_{k \ne j} \frac{Z_j(t)-Z_k(t)}{\| Z_j(t)-Z_k(t) \|^2} .
\]
Hence, if mass is to remain fixed, we must have that $\rho^3 = \theta^2$:
\[
    Z_j(t) = \rho z_j(t/\rho^{3/2}) \qquad 
    \dot{Z}_j(t) = \rho^{-1/2} \dot{z}_j(t/\rho^{3/2}) .
\]
To find the perturbation direction corresponding to this dilation, we
differentiate with respect to $\rho$ at $\rho=1$:
\[
    \frac{d}{d \rho}
    \left. \left[ \begin{array}{c} 
        \rho z_j(t/\rho^{3/2}) \\
        \rho^{-1/2} \dot{z}_j(t/\rho^{3/2})
    \end{array} \right] \right|_{\rho=1}
    =
    \left[ \begin{array}{c} 
        -\frac{3}{2} \dot{z}_j + z_j \\
	-\frac{3}{2} a_j - \frac{1}{2} \dot{z}_j
    \end{array} \right] .
\]

For checking stability, we project any initial perturbation onto the
null space of $P^T$, where
\[
  P
  =
  \left[ \begin{array}{cccccc}
      e_1 & e_2 &  0  &  0  & R z_1 & (-3\dot{z}_1+2z_1)/2 \\
      e_1 & e_2 &  0  &  0  & R z_2 & (-3\dot{z}_2+2z_2)/2 \\
      e_1 & e_2 &  0  &  0  & R z_3 & (-3\dot{z}_3+2z_3)/2 \\
       0  &  0  & e_1 & e_2 & R \dot{z}_1 & (-3a_1- \dot{z}_1)/2 \\
       0  &  0  & e_1 & e_2 & R \dot{z}_2 & (-3a_2- \dot{z}_2)/2 \\
       0  &  0  & e_1 & e_2 & R \dot{z}_3 & (-3a_3- \dot{z}_3)/2 
  \end{array} \right] .
\]
The projection matrix is given by
\[
    \Pi = I - P(P^T P)^{-1} P^T .
\]
From the fact that $z_1 + z_2 + z_3 =0$ 
and $\dot{z}_1 + \dot{z}_2 + \dot{z}_3 =0$, it follows
that all columns of $P$ are mutually orthogonal {\em except} for the
5-th and 6-th columns.  Hence, $P^T P$ is not a purely diagonal matrix.

Let
\[
    \Lambda_n
    = 
    \left(
    \prod_{i=0}^{n-1} \left( I + h A'(\xi^*(t_i)) \right)
    \right) .
\]
We say that an orbit is {\em stable} if all eigenvalues of 
\[
    \lim_{n \rightarrow \infty}
    \Lambda_n \Pi 
\]
are at most one in magnitude.

\subsection{Stable Orbits}

We computed $\Lambda_n$ for $n=10^6$.  
Table \ref{tab1} shows maximum
eigenvalues for those orbits that seemed stable from simulation.
Table \ref{tab2} shows maximum
eigenvalues for those orbits that appeared unstable when simulated.

{\bf Acknowledgements.}
The author received support from the NSF (CCR-0098040) and
the ONR (N00014-98-1-0036).

\bibliography{../../lib/refs}   

\begin{thebibliography}{8}
\expandafter\ifx\csname natexlab\endcsname\relax\def\natexlab#1{#1}\fi
\expandafter\ifx\csname url\endcsname\relax
  \def\url#1{{\tt #1}}\fi

\bibitem[Broucke(2003)]{Bro03}
R.~Broucke.
\newblock New orbits for the $n$-body problem.
\newblock In {\em Proceedings of Conference on New Trends in Astrodynamics and
  Applications}, 2003.

\bibitem[Chenciner et~al.(2001)Chenciner, Gerver, Montgomery, and
  Sim\'{o}]{CGMS01}
A.~Chenciner, J.~Gerver, R.~Montgomery, and C.~Sim\'{o}.
\newblock Simple choreographic motions on $n$ bodies: a preliminary study.
\newblock In {\em Geometry, Mechanics and Dynamics}, 2001.

\bibitem[Chenciner and Montgomery(2000)]{CM00}
A.~Chenciner and R.~Montgomery.
\newblock A remarkable periodic solution of the three-body problem in the case
  of equal masses.
\newblock {\em Annals of Math}, 152:\penalty0 881--901, 2000.

\bibitem[Fourer et~al.(1993)Fourer, Gay, and Kernighan]{FGK93}
R.~Fourer, D.M. Gay, and B.W. Kernighan.
\newblock {\em AMPL: A Modeling Language for Mathematical Programming}.
\newblock Scientific Press, 1993.

\bibitem[Moore(1993)]{Mor93}
C.~Moore.
\newblock Braids in classical gravity.
\newblock {\em Phys. Rev. Lett.}, 70:\penalty0 3675--3679, 1993.

\bibitem[Vanderbei(1999)]{SOR9708}
R.J. Vanderbei.
\newblock {L}{O}{Q}{O} user's manual---version 3.10.
\newblock {\em Optimization Methods and Software}, 12:\penalty0 485--514, 1999.

\bibitem[Vanderbei(2001)]{GravityApplet}
R.J. Vanderbei.
\newblock
  http://www.princeton.edu/$\sim$rvdb/{J}{A}{V}{A}/astro/galaxy/{G}alaxy.html,
  2001.
\newblock ~.

\bibitem[Vanderbei and Shanno(1999)]{Van97d}
R.J. Vanderbei and D.F. Shanno.
\newblock An interior-point algorithm for nonconvex nonlinear programming.
\newblock {\em Computational Optimization and Applications}, 13:\penalty0
  231--252, 1999.

\end{thebibliography}
\bibliographystyle{plainnat}   

\clearpage

\begin{table}
\begin{center}
\begin{tabular}{rrr}
Name & $\max(\lambda_i(\Lambda))$ & $\max(\lambda_i(\Lambda \Pi))$ \\
\hline
Lagrange2 & 1.383 & 1.362 \\
FigureEight3 & 1.228 & 4.220 \\
Ducati3 & 1.105 & 3.885 \\
Hill3\_15 & 1.444 & 2.403 \\
DoubleDouble5 & 12.298 & 12.298 \\
DoubleDouble10 & 1.404 & 5.948 \\
DoubleDouble20 & 1.890 & 1.890 
\end{tabular}
\end{center}
\caption{Apparently stable orbits.}
\label{tab1}
\end{table}

\begin{table}
\begin{center}
\begin{tabular}{rrr}
Name & $\max(\lambda_i(\Lambda))$ & $\max(\lambda_i(\Lambda \Pi))$ \\
\hline
Lagrange3 & 81.630 & 81.630 \\
OrthQuasiEllipse4 & 18.343 & 18.343 \\
Rosette4 & 1.873 & 4.449 \\
Braid4 & 727.508 & 711.811 \\
Trefoil4 & 41228.515 & 41213.852 \\
FigureEight4 & 221.642 & 194.095 \\
FoldedTriLoop4 & 74758.355 & 74675.092 \\
PlateSaucer4 & 3653.210 & 3653.210 \\
BorderCollie4 & 188.235 & 188.052 \\
Trefoil5 & 1.913e+8 & 1.917e+8 \\
FigureEight5 &  2223.137 & 2223.457 
\end{tabular}
\end{center}
\caption{Apparently unstable orbits.}
\label{tab2}
\end{table}

\begin{figure}
\scriptsize
\begin{verbatim}
param N := 3;  # number of masses
param n := 15;  # number of terms in Fourier series representation
param m := 100;  # number of terms in numerical approx to integral

set Bodies := {0..N-1};
set Times  := {0..m-1} circular; # "circular" means that next(m-1) = 0

param theta {t in Times} := t*2*pi/m;
param dt := 2*pi/m;

param a0 {i in Bodies} default 0;        param b0 {i in Bodies} default 0;
var as {i in Bodies, k in 1..n} := 0;    var bs {i in Bodies, k in 1..n} := 0;
var ac {i in Bodies, k in 1..n} := 0;    var bc {i in Bodies, k in 1..n} := 0;

var x {i in Bodies, t in Times} 
  = a0[i]+sum {k in 1..n} ( as[i,k]*sin(k*theta[t]) + ac[i,k]*cos(k*theta[t]) );
var y {i in Bodies, t in Times} 
  = b0[i]+sum {k in 1..n} ( bs[i,k]*sin(k*theta[t]) + bc[i,k]*cos(k*theta[t]) );

var xdot {i in Bodies, t in Times} = (x[i,next(t)]-x[i,t])/dt;
var ydot {i in Bodies, t in Times} = (y[i,next(t)]-y[i,t])/dt;

var K {t in Times} = 0.5*sum {i in Bodies} (xdot[i,t]^2 + ydot[i,t]^2);

var P {t in Times}
  = - sum {i in Bodies, ii in Bodies: ii>i} 
            1/sqrt((x[i,t]-x[ii,t])^2 + (y[i,t]-y[ii,t])^2);

minimize A: sum {t in Times} (K[t] - P[t])*dt;

let {i in Bodies, k in 1..n} as[i,k] := 1*(Uniform01()-0.5);
let {i in Bodies, k in 1..n} ac[i,k] := 1*(Uniform01()-0.5);
let {i in Bodies, k in n..n} bs[i,k] := 0.01*(Uniform01()-0.5);
let {i in Bodies, k in n..n} bc[i,k] := 0.01*(Uniform01()-0.5);

solve;
\end{verbatim}
\caption{{\sc ampl} program for finding trajectories that minimize the action
functional.} \label{fig:50}
\end{figure}

\begin{figure}
\scriptsize
\begin{verbatim}
param N := 3;  # number of masses
param n := 15;  # number of terms in Fourier series representation
param m := 99;  # terms in num approx to integral.  must be a multiple of N

param lagTime := m/N;

set Bodies := {0..N-1};
set Times  := {0..m-1} circular; # "circular" means that next(m-1) = 0

param theta {t in Times} := t*2*pi/m;
param dt := 2*pi/m;

param a0 default 0;         param b0 default 0;
var as {k in 1..n} := 0;    var bs {k in 1..n} := 0;
var ac {k in 1..n} := 0;    var bc {k in 1..n} := 0;

var x {i in Bodies, t in Times} 
  = a0+sum {k in 1..n} ( as[k]*sin(k*theta[(t+i*lagTime) mod m]) 
                       + ac[k]*cos(k*theta[(t+i*lagTime) mod m]) );
var y {i in Bodies, t in Times} 
  = b0+sum {k in 1..n} ( bs[k]*sin(k*theta[(t+i*lagTime) mod m]) 
                       + bc[k]*cos(k*theta[(t+i*lagTime) mod m]) );

var xdot {i in Bodies, t in Times} = (x[i,next(t)]-x[i,t])/dt;
var ydot {i in Bodies, t in Times} = (y[i,next(t)]-y[i,t])/dt;

var K {t in Times} = 0.5*sum {i in Bodies} (xdot[i,t]^2 + ydot[i,t]^2);

var P {t in Times}
  = - sum {i in Bodies, ii in Bodies: ii>i} 
            1/sqrt((x[i,t]-x[ii,t])^2 + (y[i,t]-y[ii,t])^2);

minimize A: sum {t in Times} (K[t] - P[t])*dt;

let {k in 1..n} as[k] := 1*(Uniform01()-0.5);
let {k in 1..n} ac[k] := 1*(Uniform01()-0.5);
let {k in n..n} bs[k] := 0.01*(Uniform01()-0.5);
let {k in n..n} bc[k] := 0.01*(Uniform01()-0.5);

solve;
\end{verbatim}
\caption{{\sc ampl} program for finding choreographies by minimizing the action
functional.} \label{fig:51}
\end{figure}

\begin{figure}
\begin{center} 

\begin{minipage}{2in} \begin{center}
\includegraphics[width=2in]{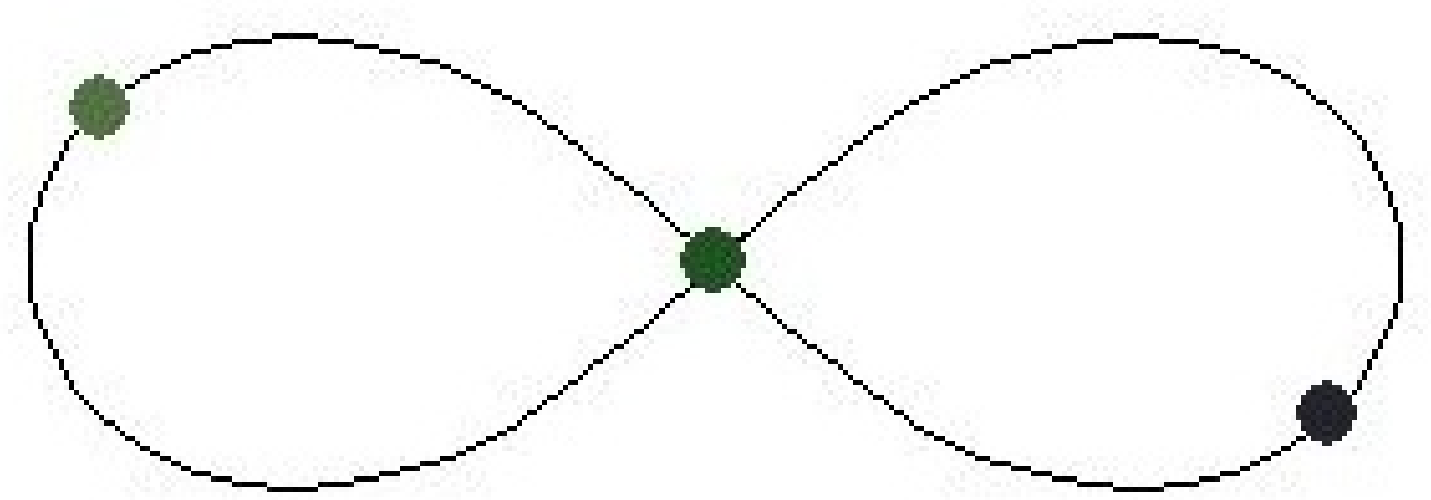} \\
FigureEight3
\end{center}\end{minipage}
\begin{minipage}{2in} \begin{center}
\includegraphics[width=2in]{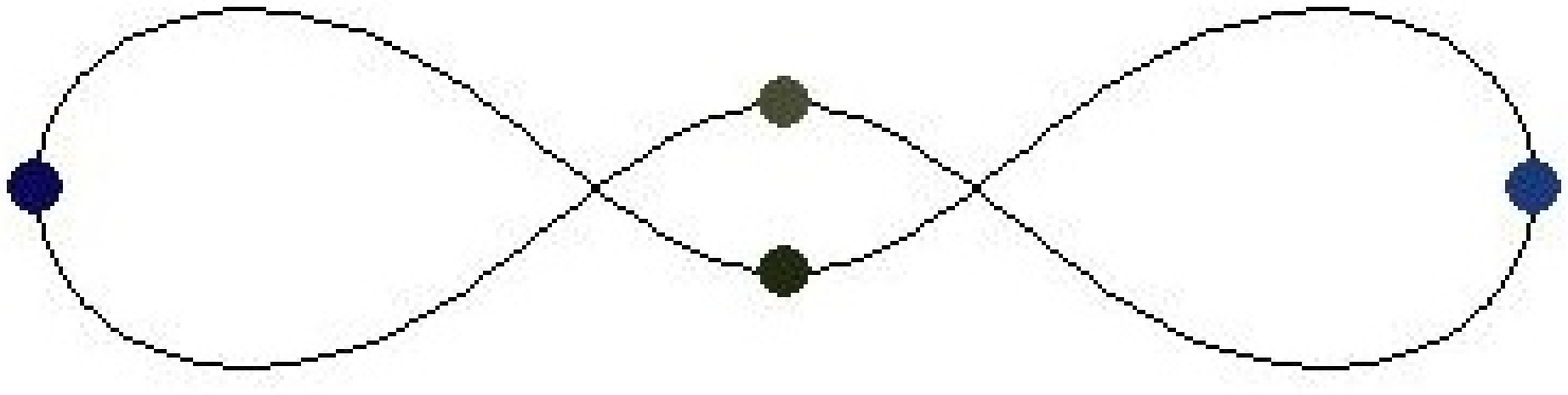} \\
Braid4
\end{center}\end{minipage}
\begin{minipage}{2in} \begin{center}
\includegraphics[width=2in]{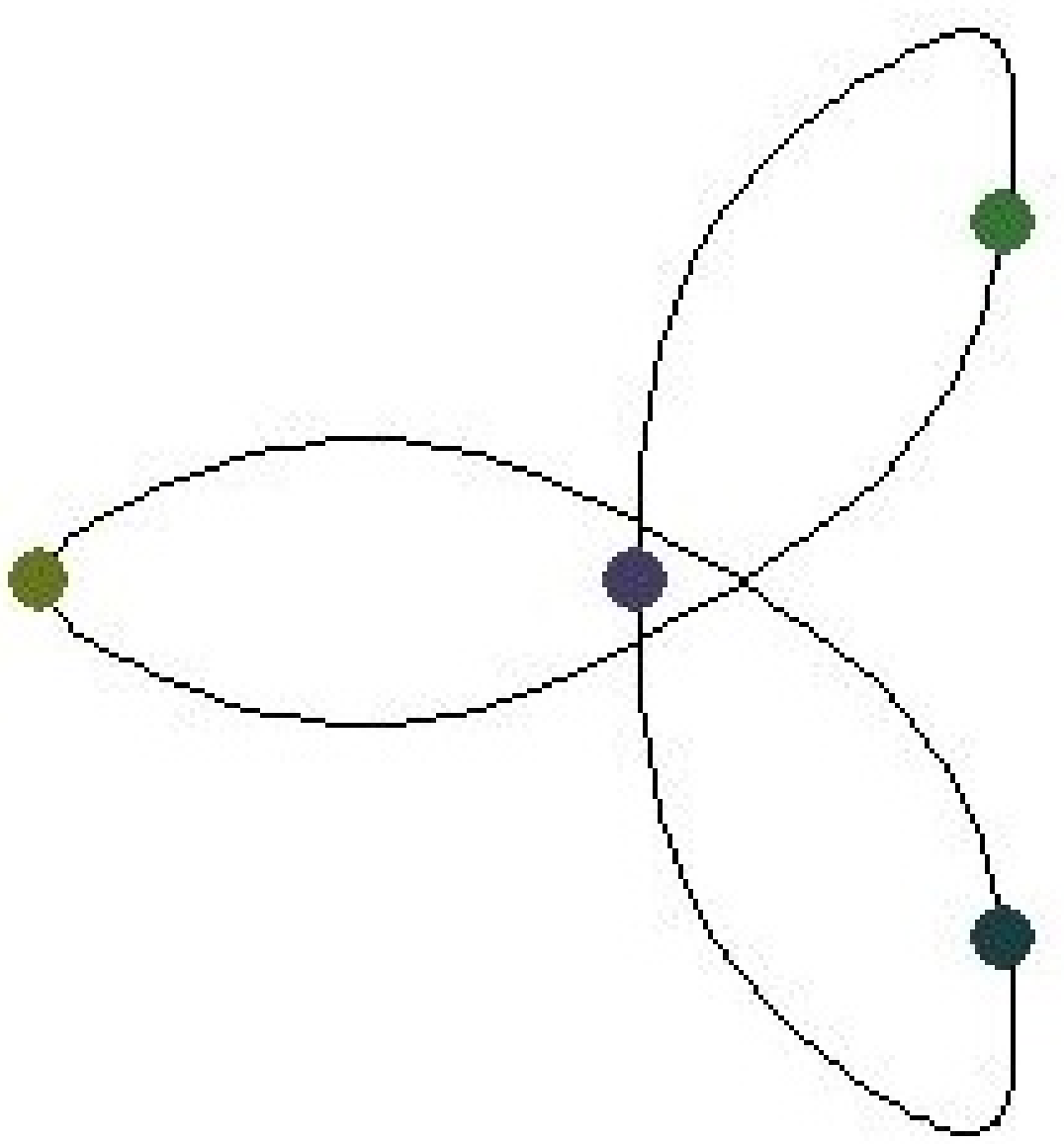} \\
Trefoil4
\end{center}\end{minipage}
\begin{minipage}{2in} \begin{center}
\includegraphics[width=2in]{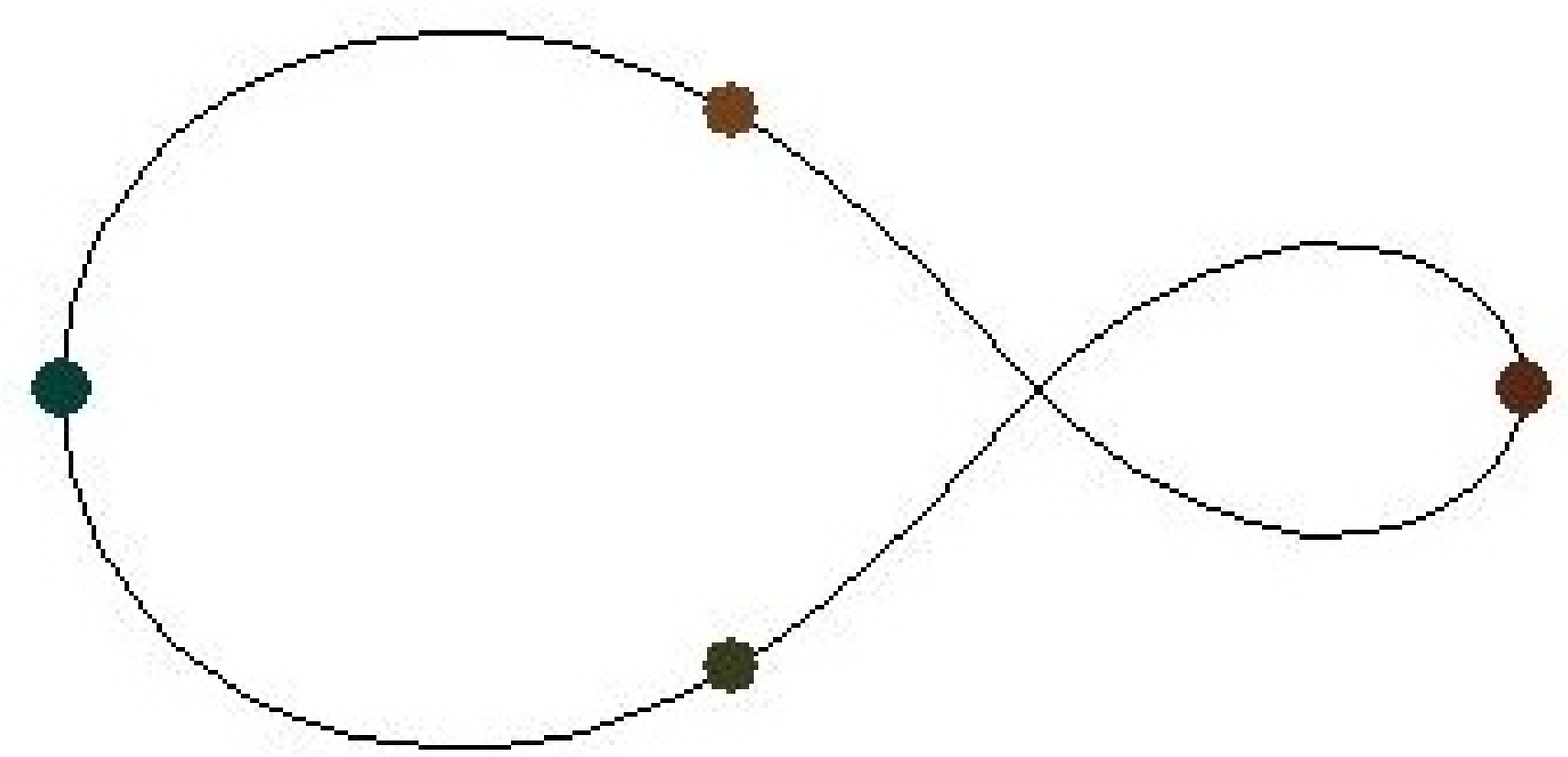} \\
FigureEight4
\end{center}\end{minipage}
\begin{minipage}{2in} \begin{center}
\includegraphics[width=2in]{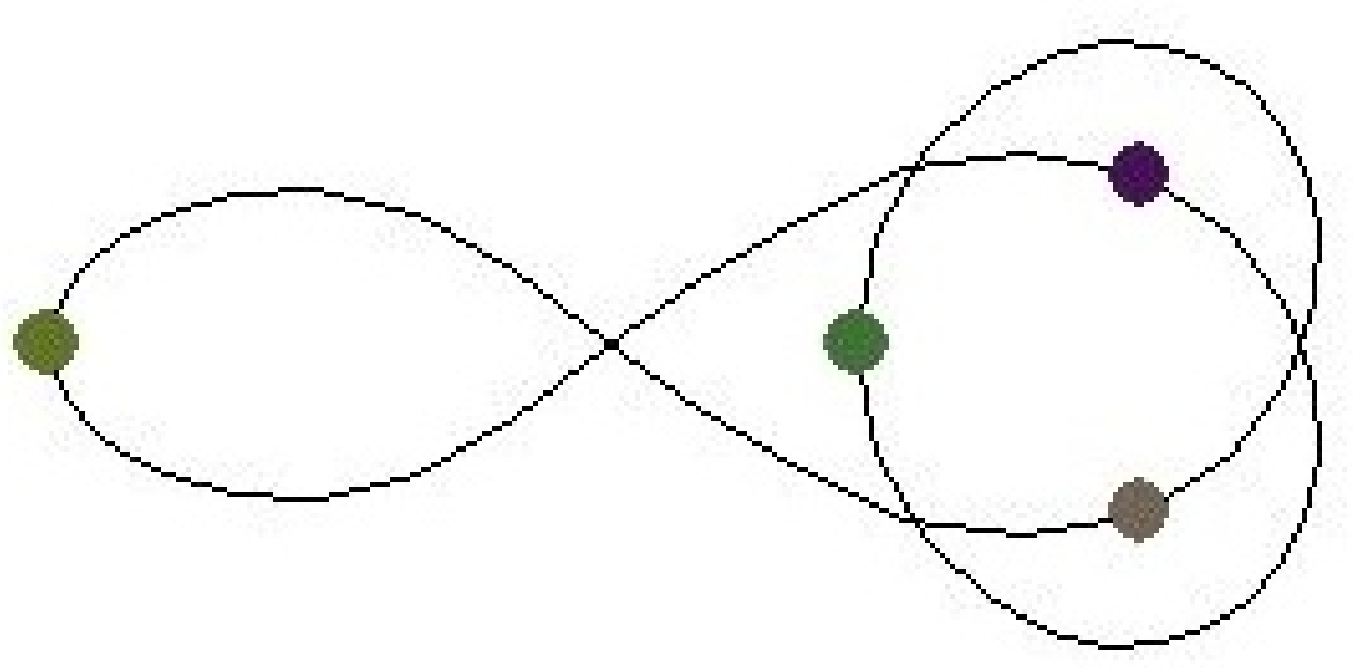} \\
FoldedTriLoop4
\end{center}\end{minipage}
\begin{minipage}{2in} \begin{center}
\includegraphics[width=2in]{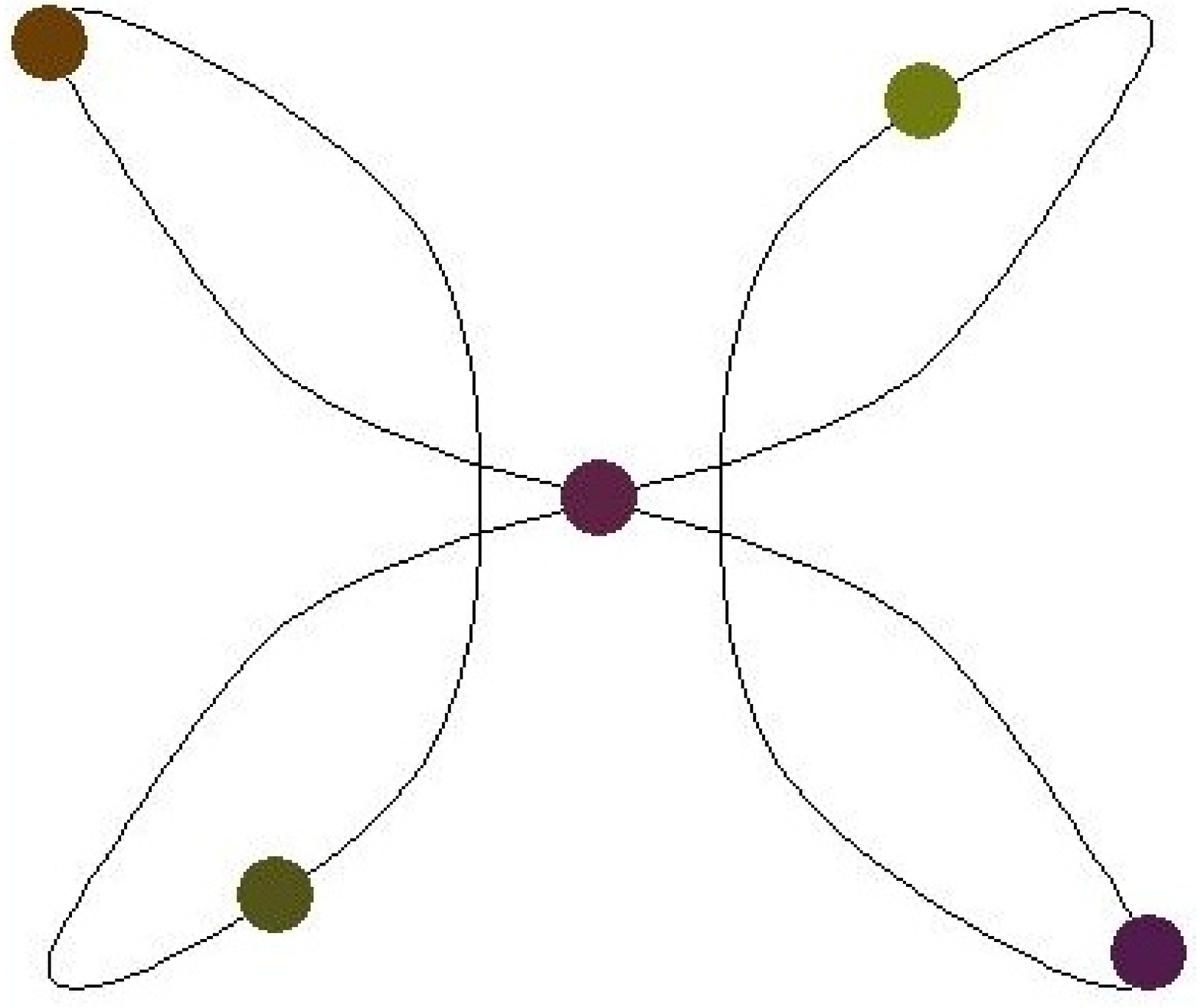} \\
Trefoil5
\end{center}\end{minipage}
\begin{minipage}{2in} \begin{center}
\includegraphics[width=2in]{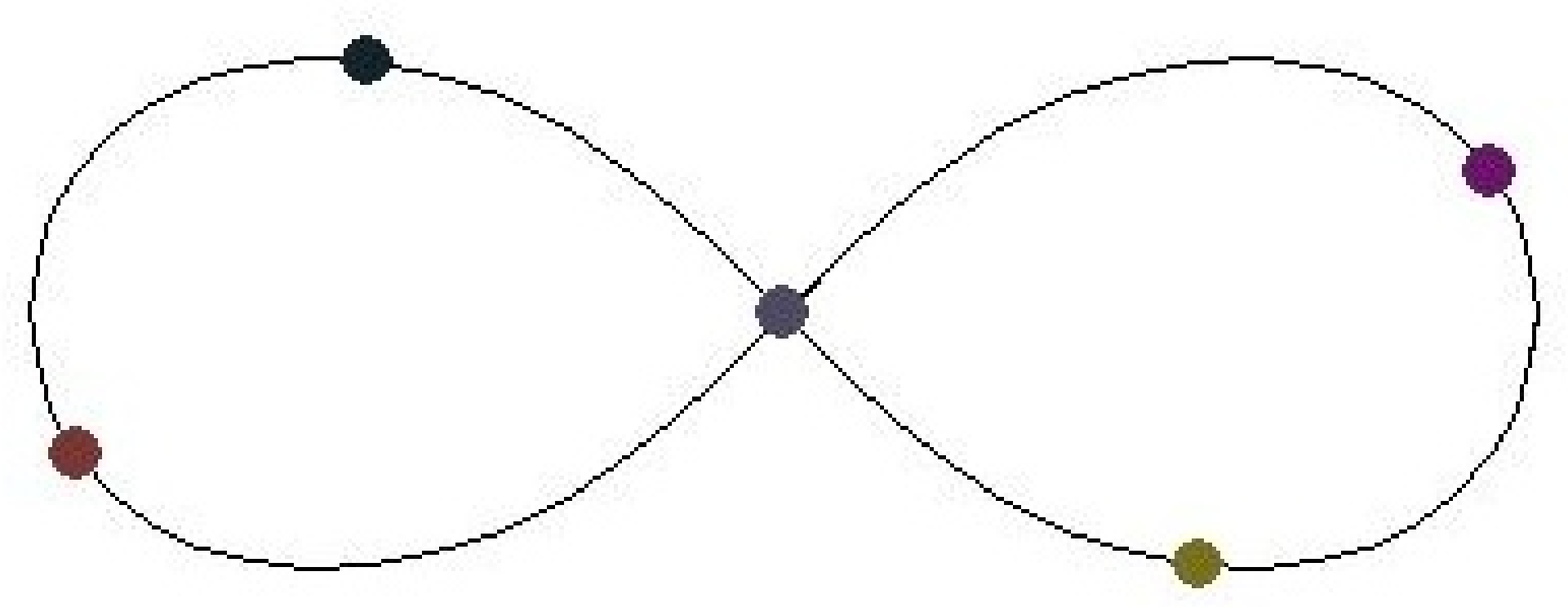} \\
FigureEight5
\end{center}\end{minipage}
\end{center}
\caption{Periodic Orbits---Choreographies.}
\label{fig:fig1}
\end{figure}

\begin{figure}
\begin{center} 

\begin{minipage}{2in} \begin{center}
\includegraphics[width=2in]{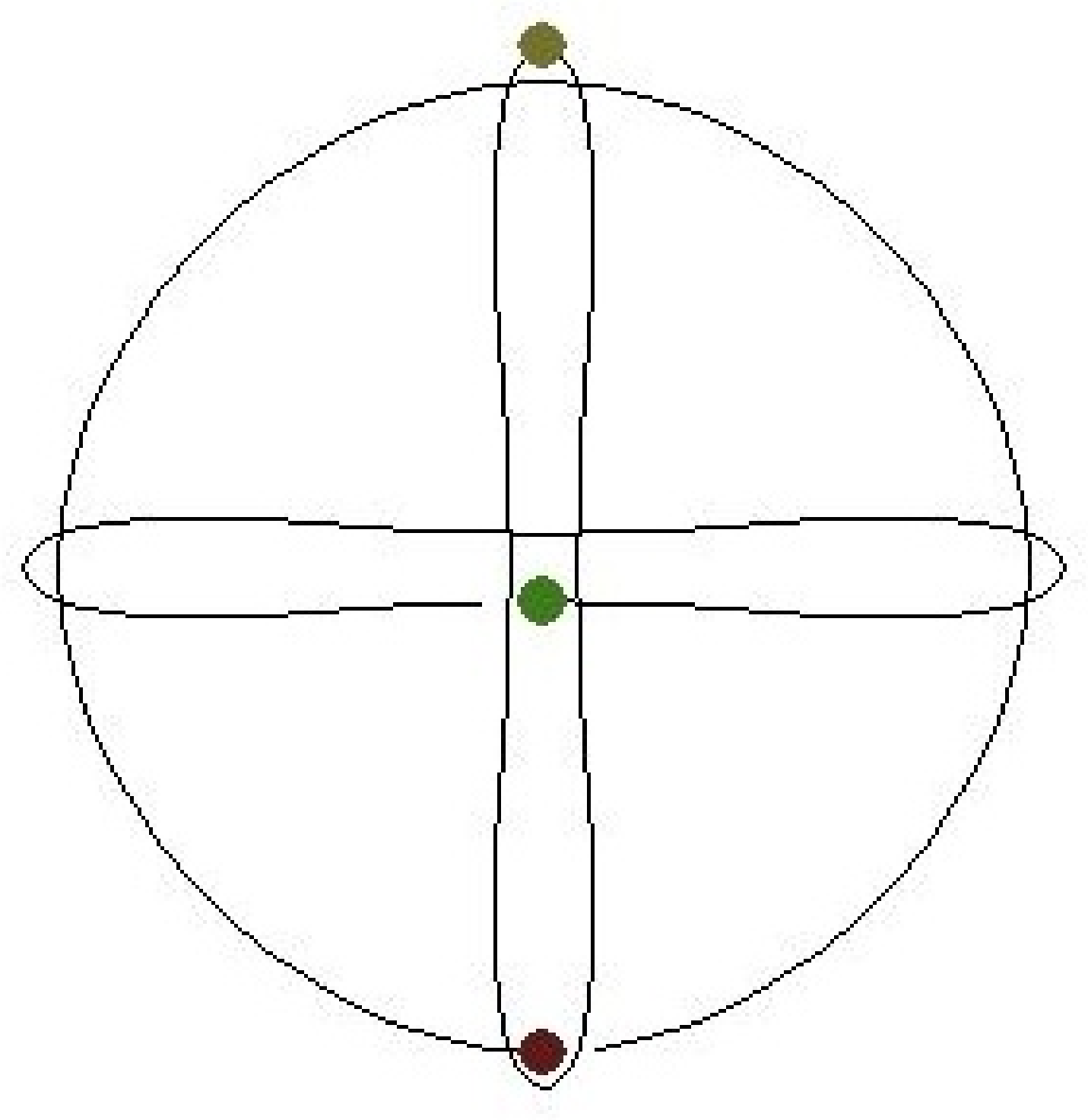} \\
Ducati3
\end{center}\end{minipage}
\begin{minipage}{2in} \begin{center}
\includegraphics[width=2in]{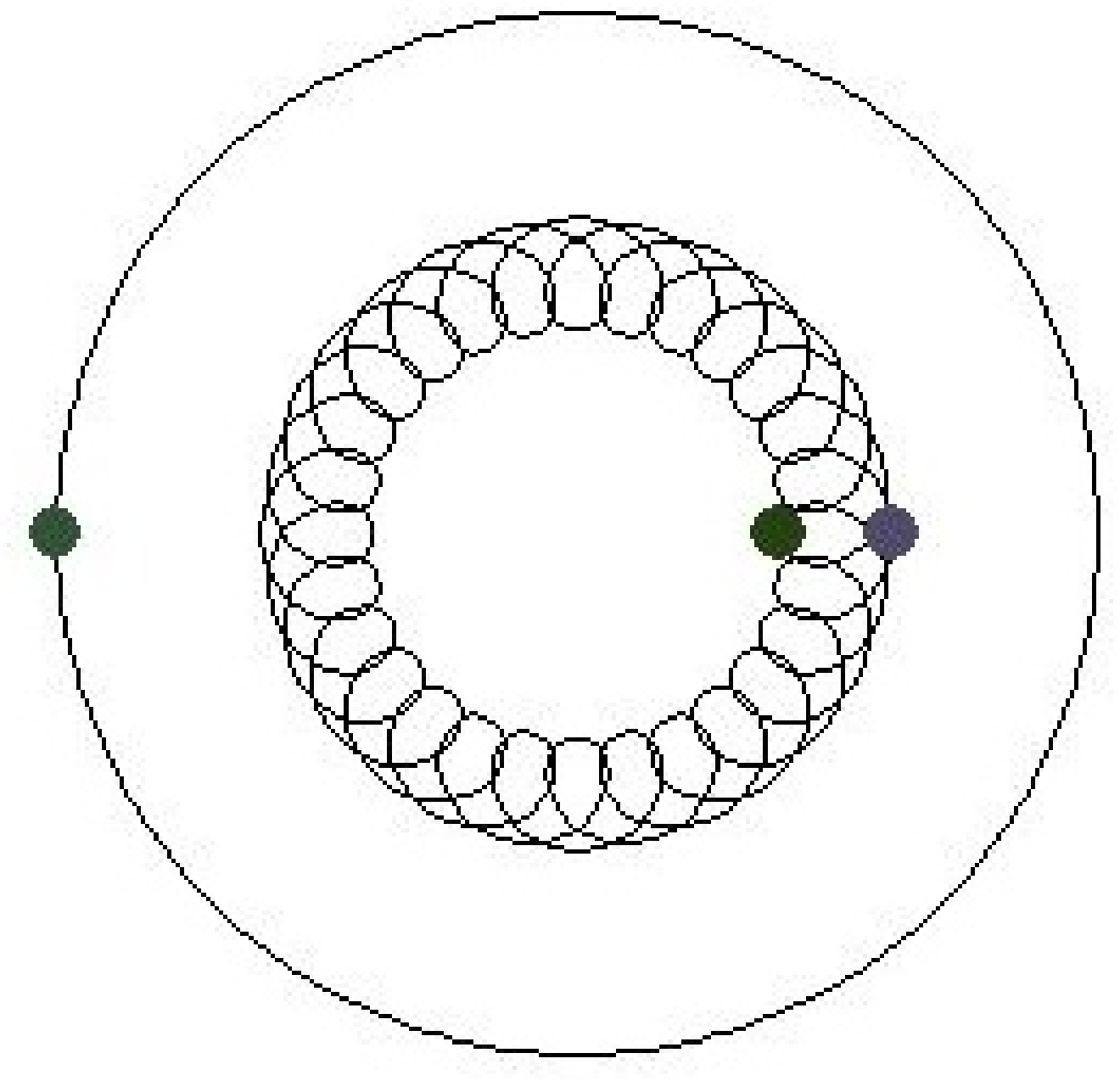} \\
Hill3\_15
\end{center}\end{minipage}
\begin{minipage}{2in} \begin{center}
\includegraphics[width=2in]{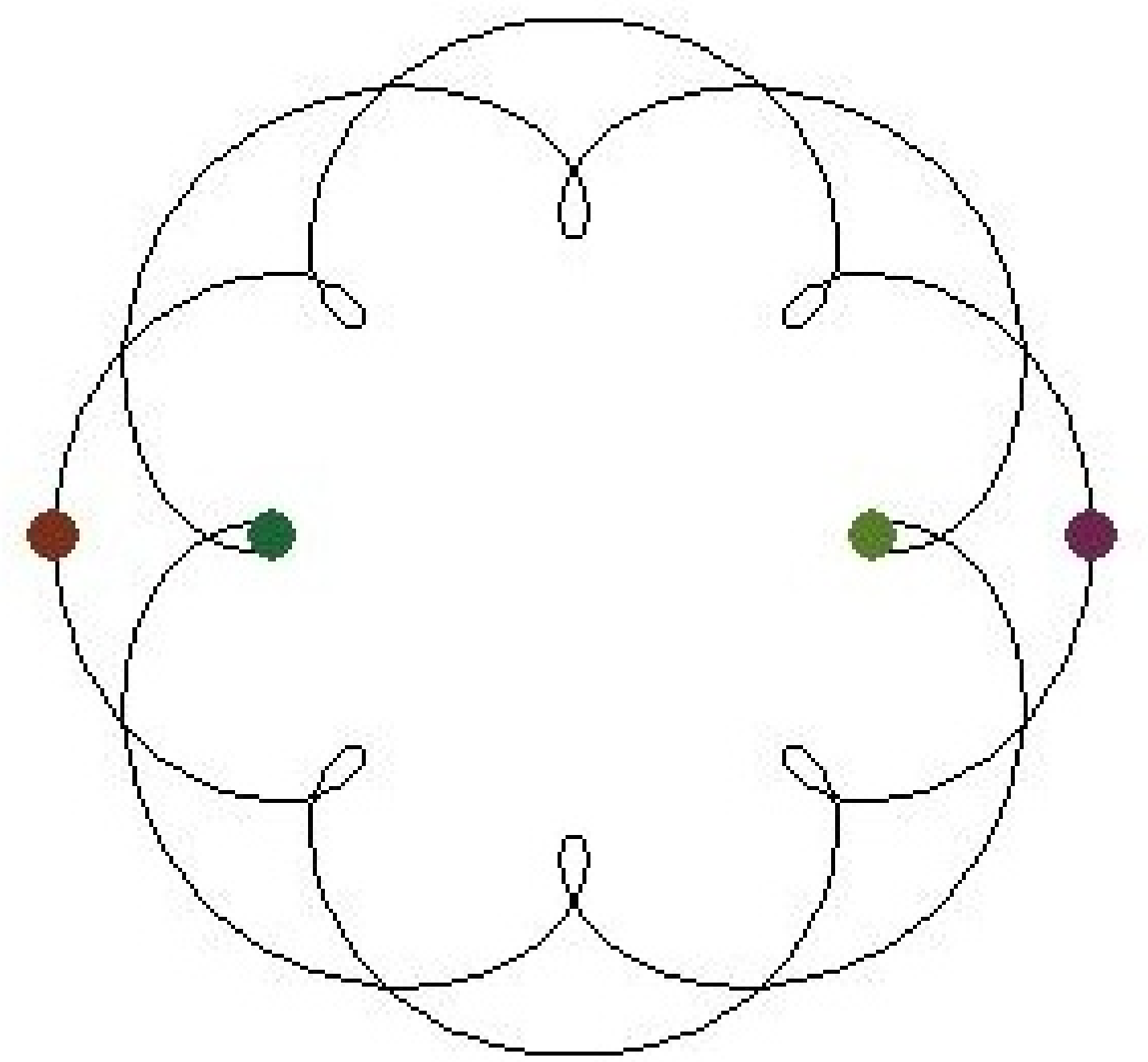} \\
DoubleDouble5
\end{center}\end{minipage}
\begin{minipage}{2in} \begin{center}
\includegraphics[width=2in]{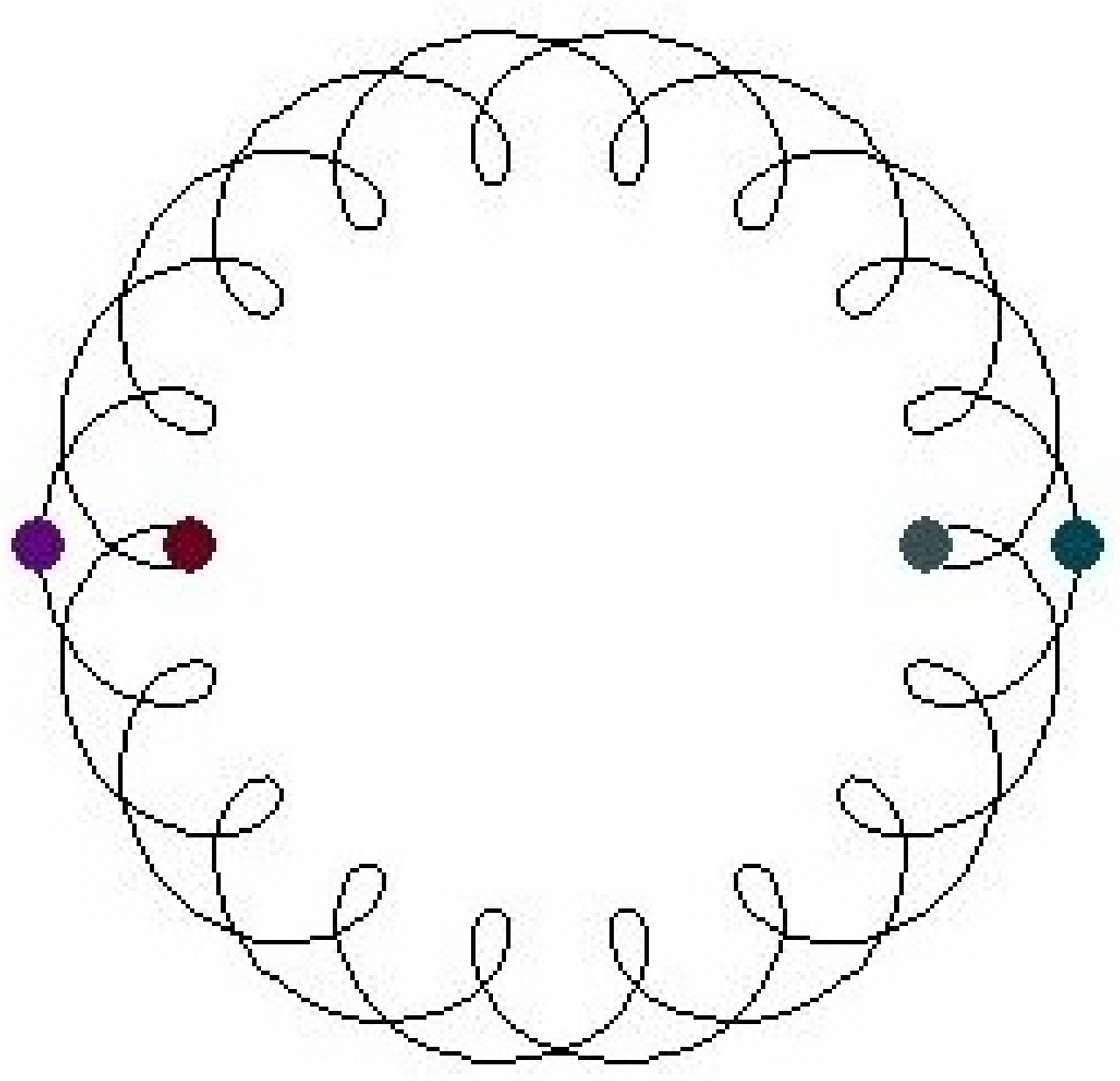} \\
DoubleDouble10
\end{center}\end{minipage}
\begin{minipage}{2in} \begin{center}
\includegraphics[width=2in]{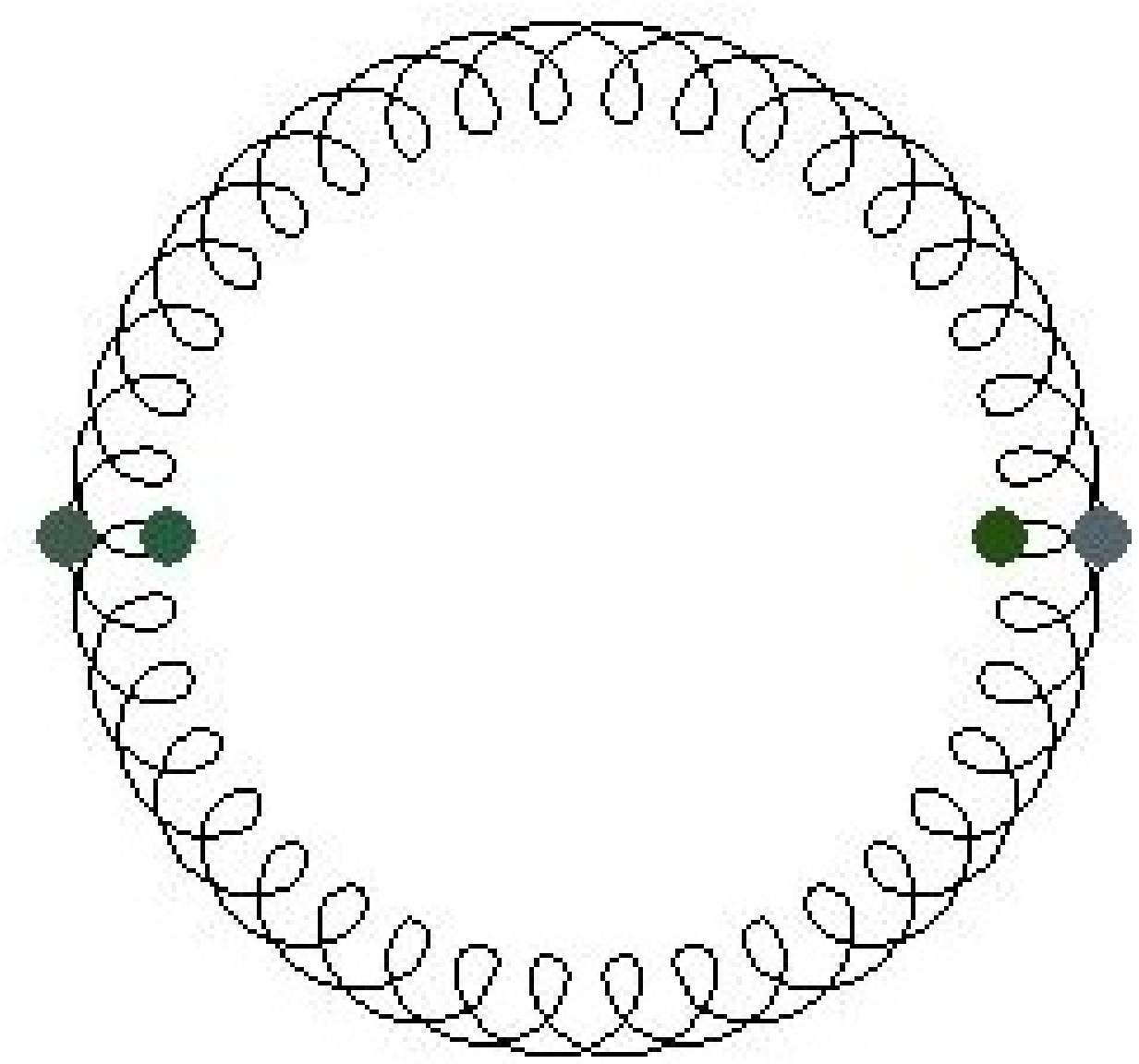} \\
DoubleDouble20
\end{center}\end{minipage}
\begin{minipage}{2in} \begin{center}
\includegraphics[width=2in]{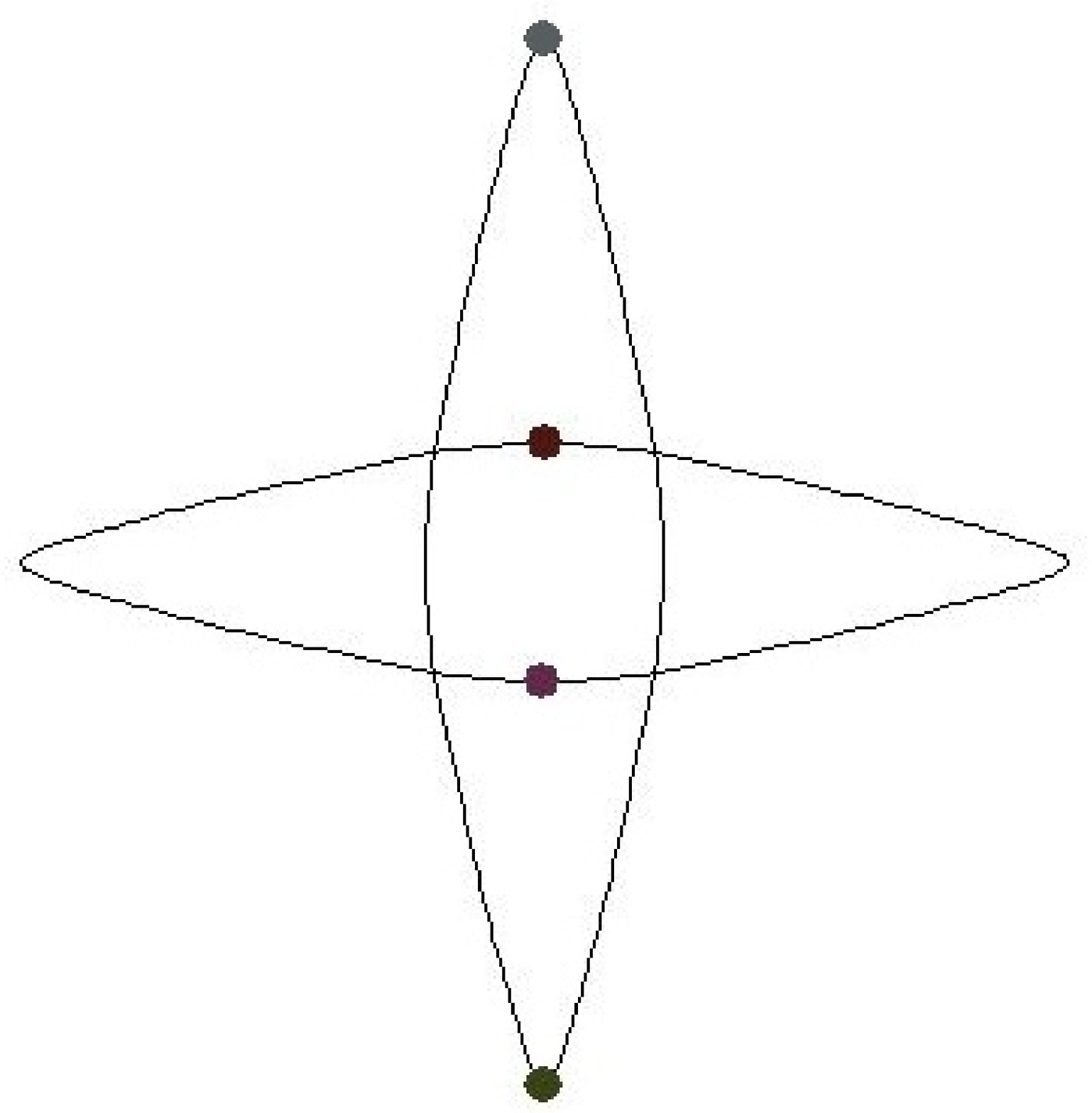} \\
OrthQuasiEllipse4
\end{center}\end{minipage}
\begin{minipage}{2in} \begin{center}
\includegraphics[width=2in]{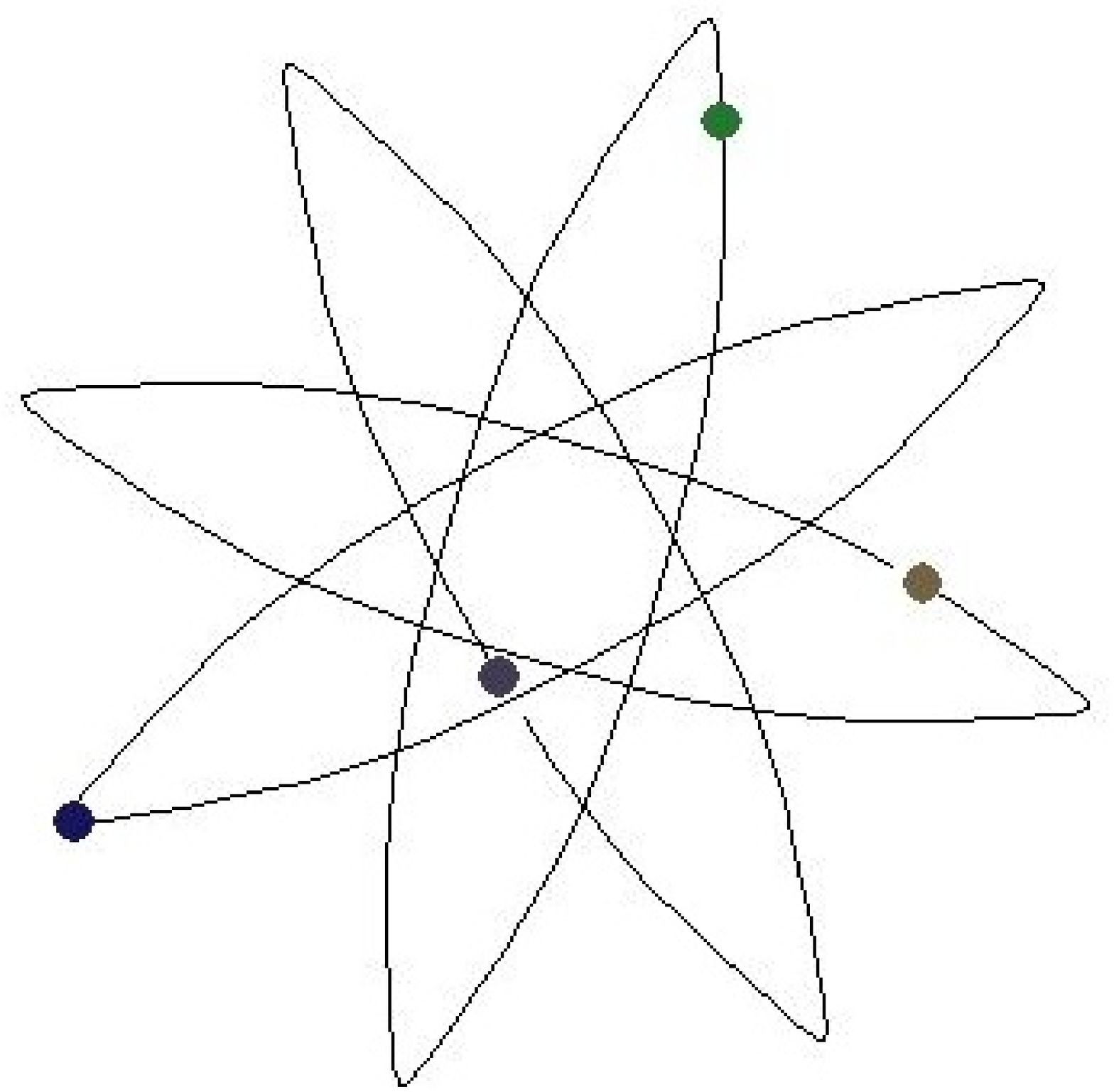} \\
Rosette4
\end{center}\end{minipage}
\begin{minipage}{2in} \begin{center}
\includegraphics[width=2in]{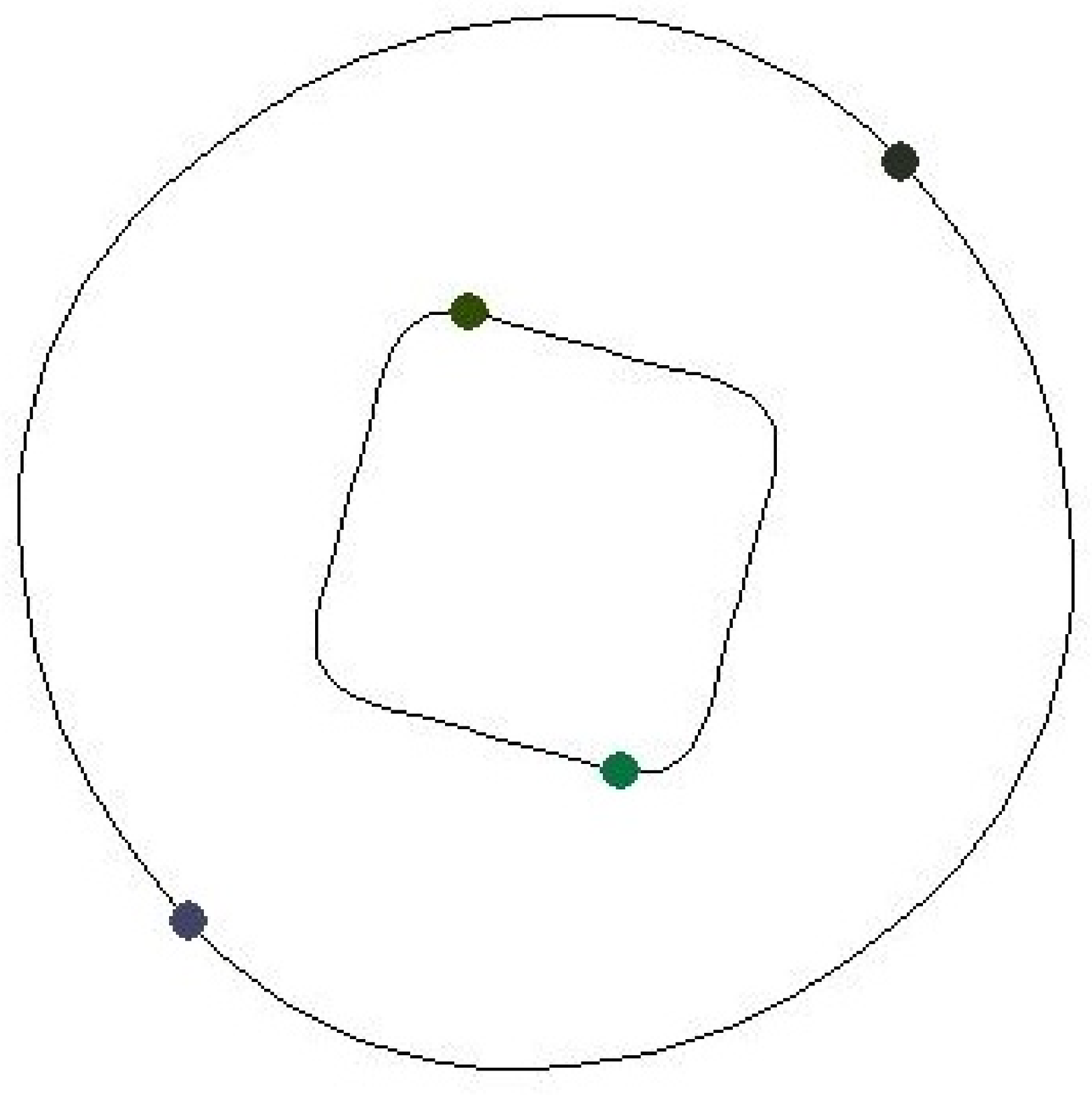} \\
PlateSaucer4
\end{center}\end{minipage}
\begin{minipage}{2in} \begin{center}
\includegraphics[width=2in]{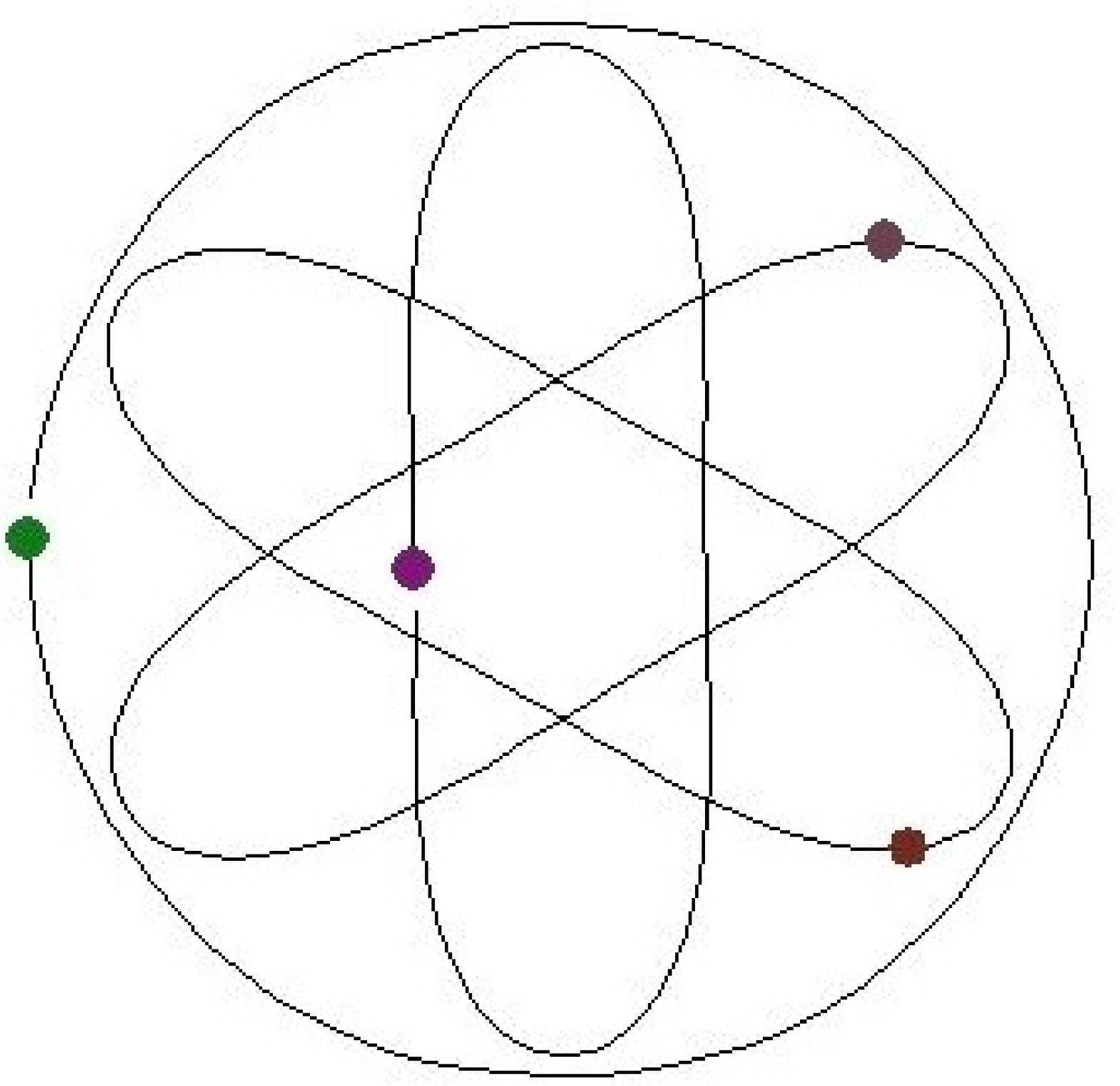} \\
BorderCollie4
\end{center}\end{minipage}
\end{center}
\caption{Periodic Orbits--Non-Choreographies.}
\label{fig:fig2}
\end{figure}

\begin{figure}
\begin{center} 

\begin{minipage}{2in} \begin{center}
\includegraphics[width=2in]{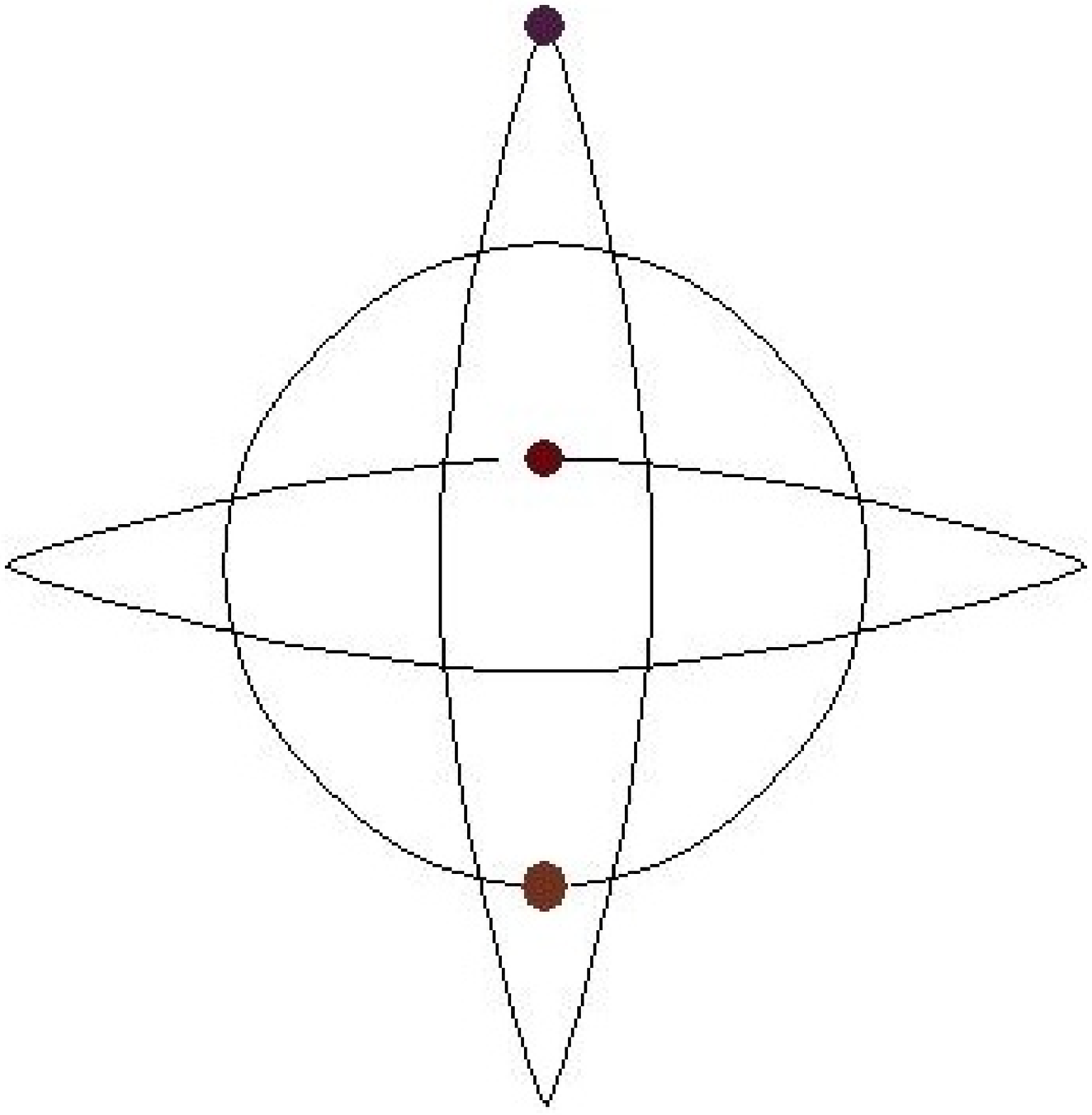} \\
Ducati3\_2
\end{center}\end{minipage}
\begin{minipage}{2in} \begin{center}
\includegraphics[width=2in]{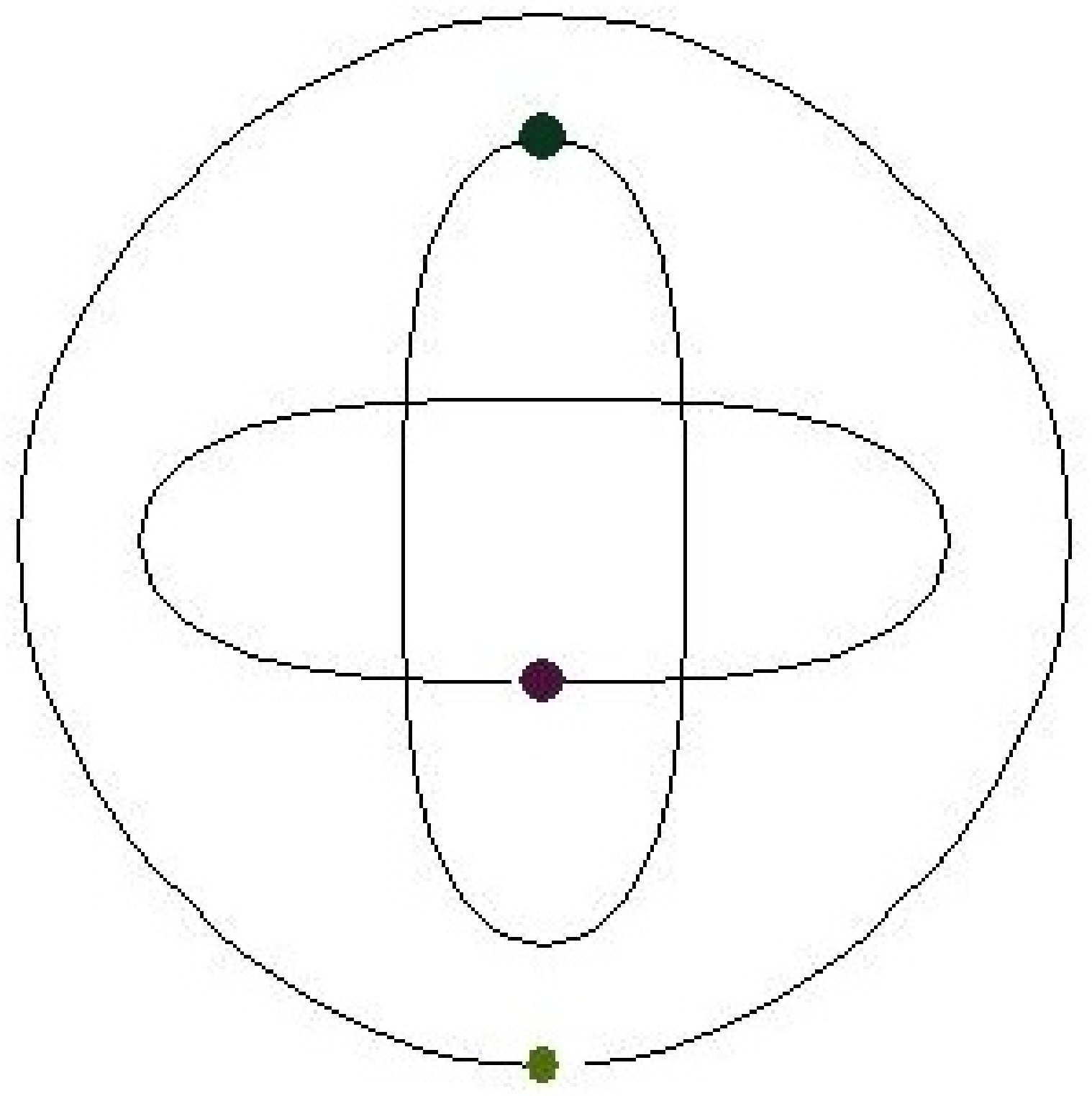} \\
Ducati3\_0.5
\end{center}\end{minipage}
\begin{minipage}{2in} \begin{center}
\includegraphics[width=2in]{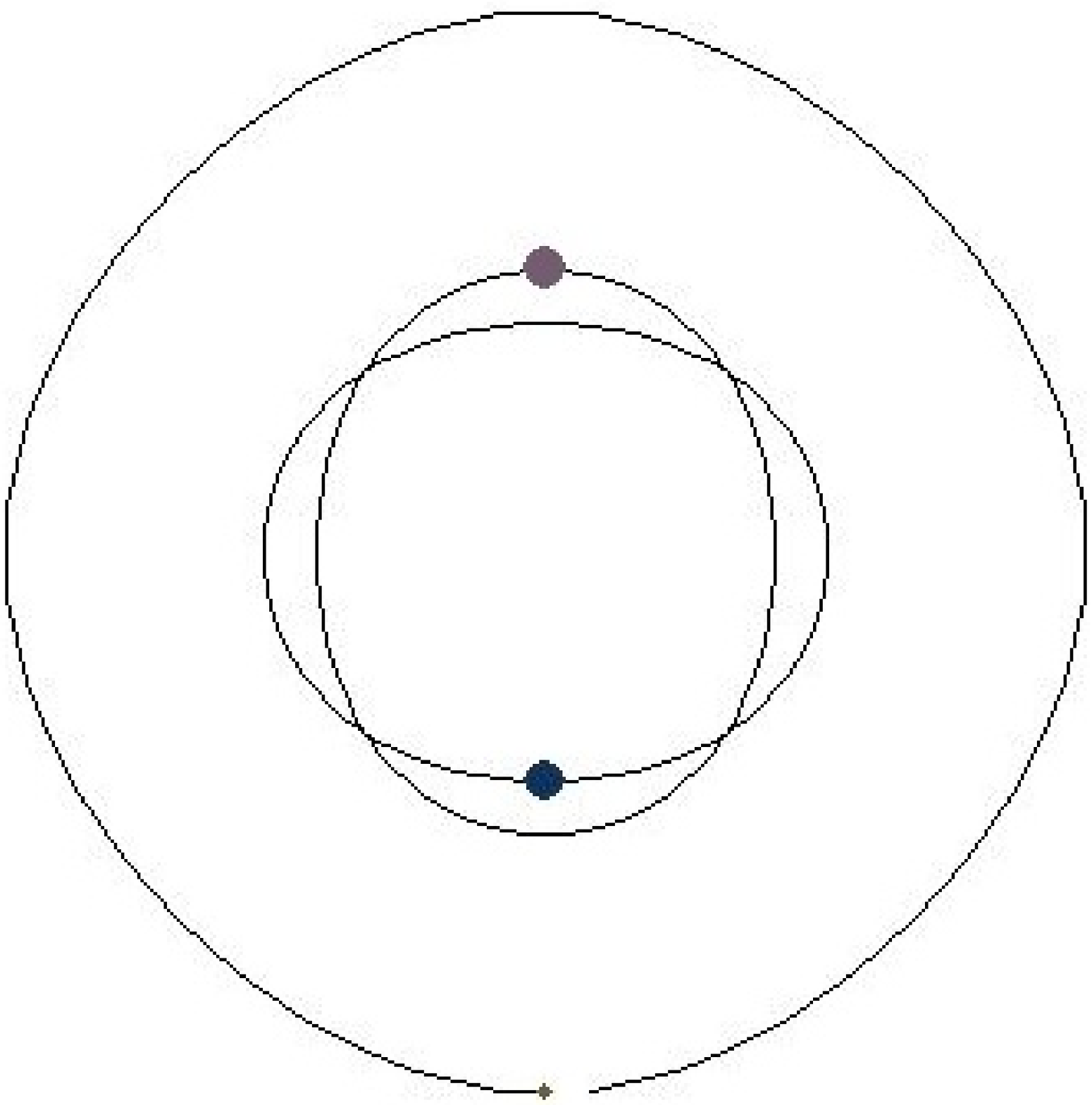} \\
Ducati3\_0.1
\end{center}\end{minipage}

\begin{minipage}{2in} \begin{center}
\includegraphics[width=2in]{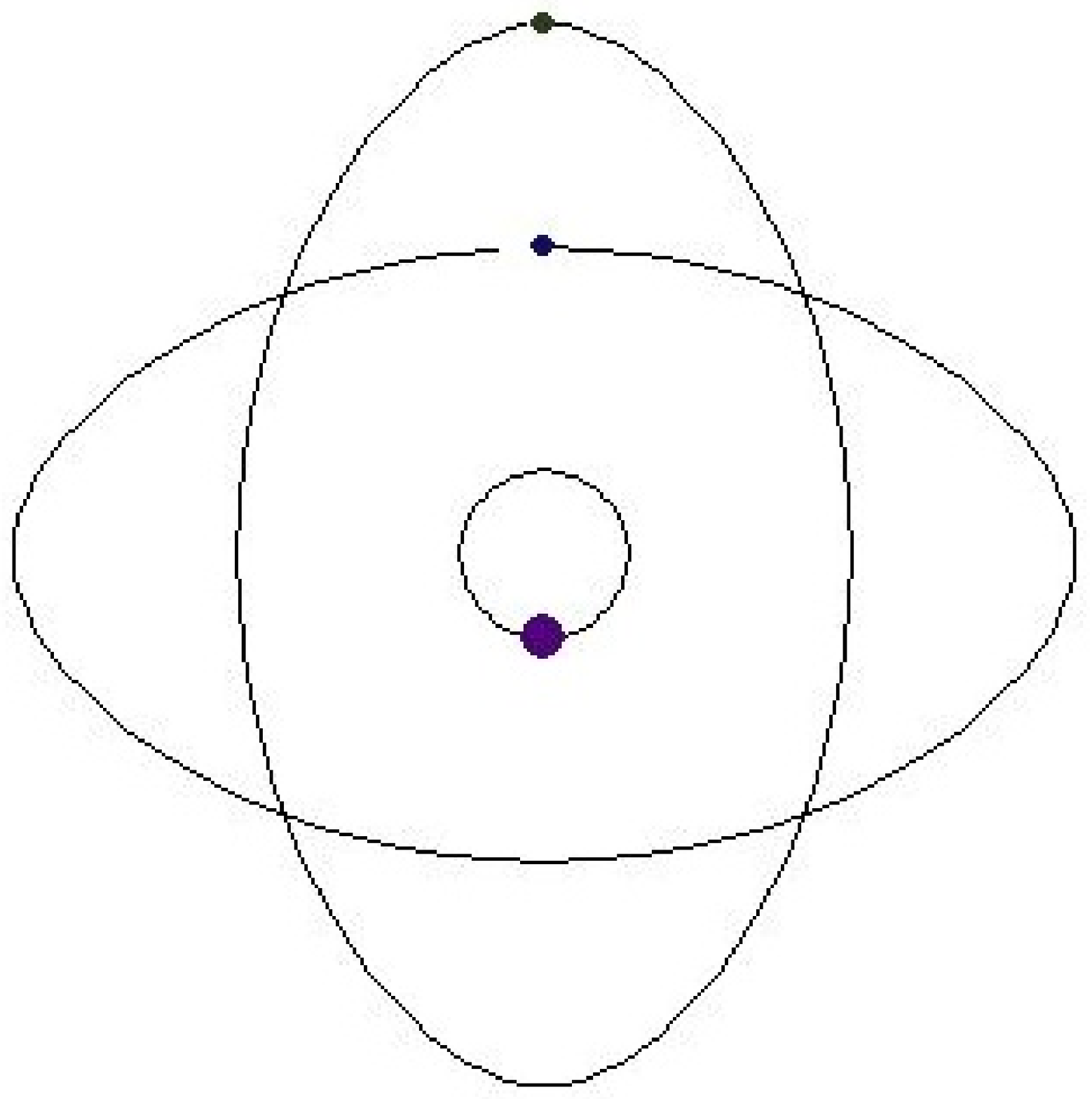} \\
Ducati3\_10
\end{center}\end{minipage}
\begin{minipage}{2in} \begin{center}
\includegraphics[width=2in]{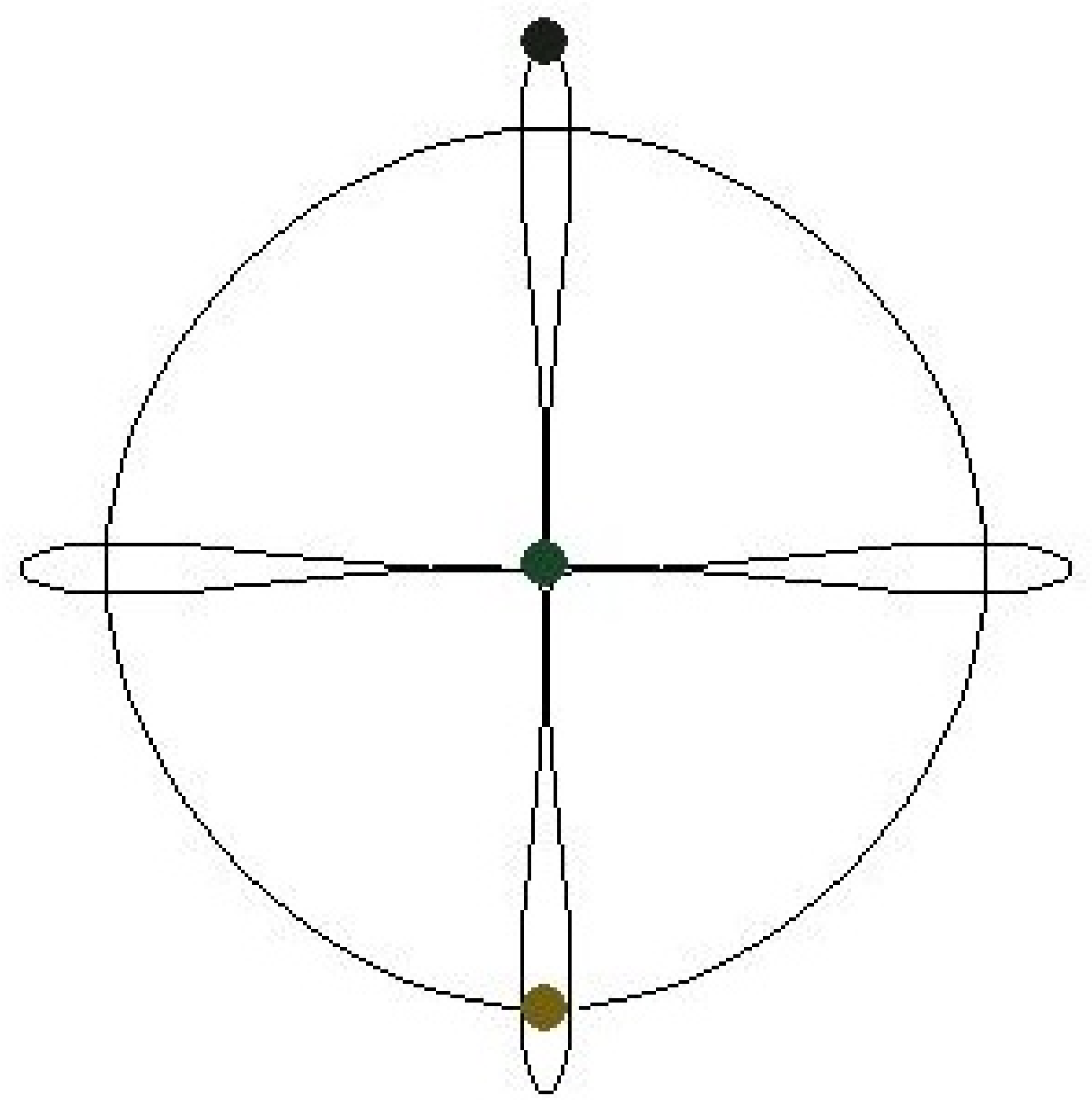} \\
Ducati3\_1.2
\end{center}\end{minipage}
\begin{minipage}{2in} \begin{center}
\includegraphics[width=2in]{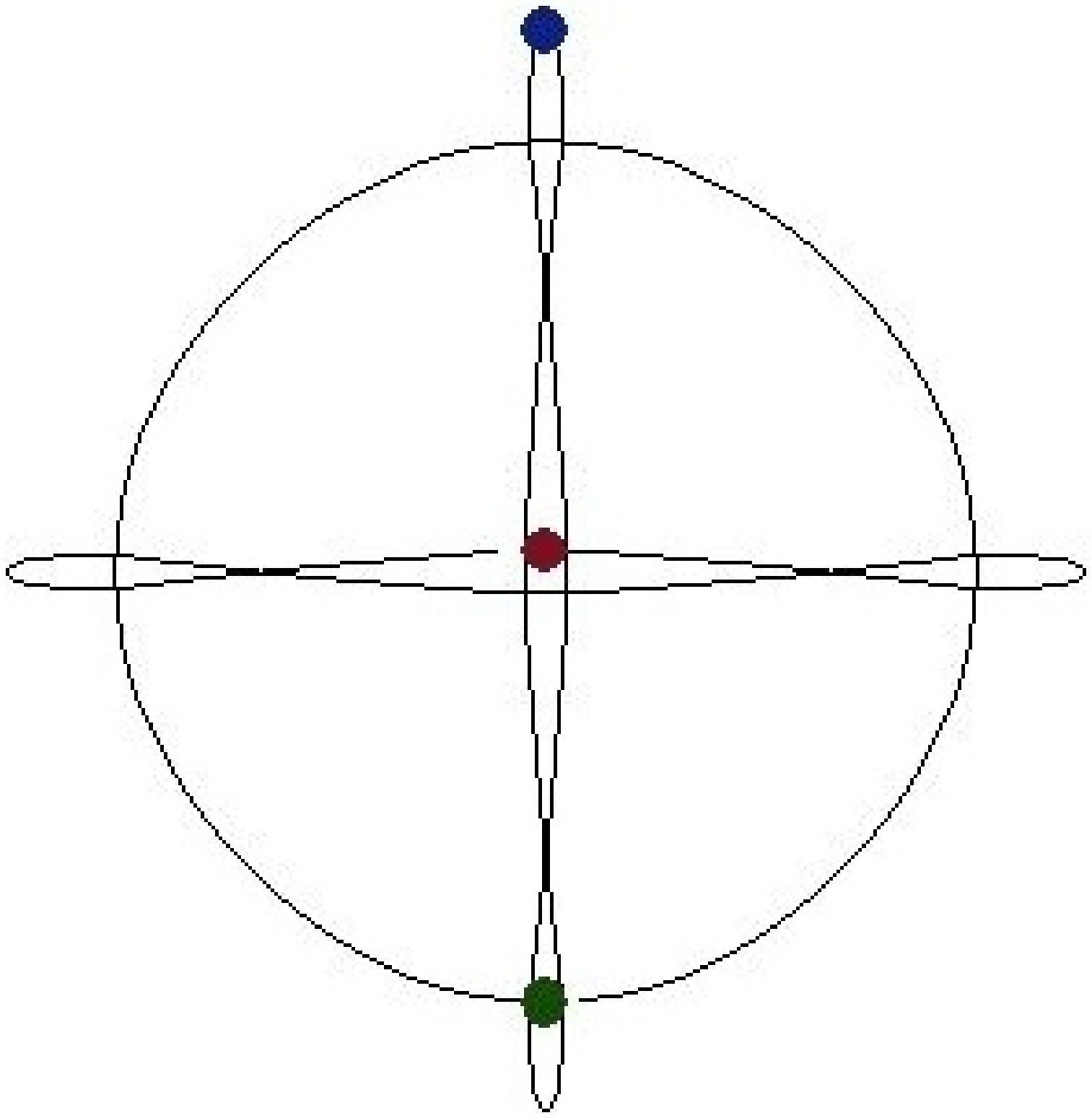} \\
Ducati3\_1.3
\end{center}\end{minipage}
\begin{minipage}{2in} \begin{center}
\includegraphics[width=2in]{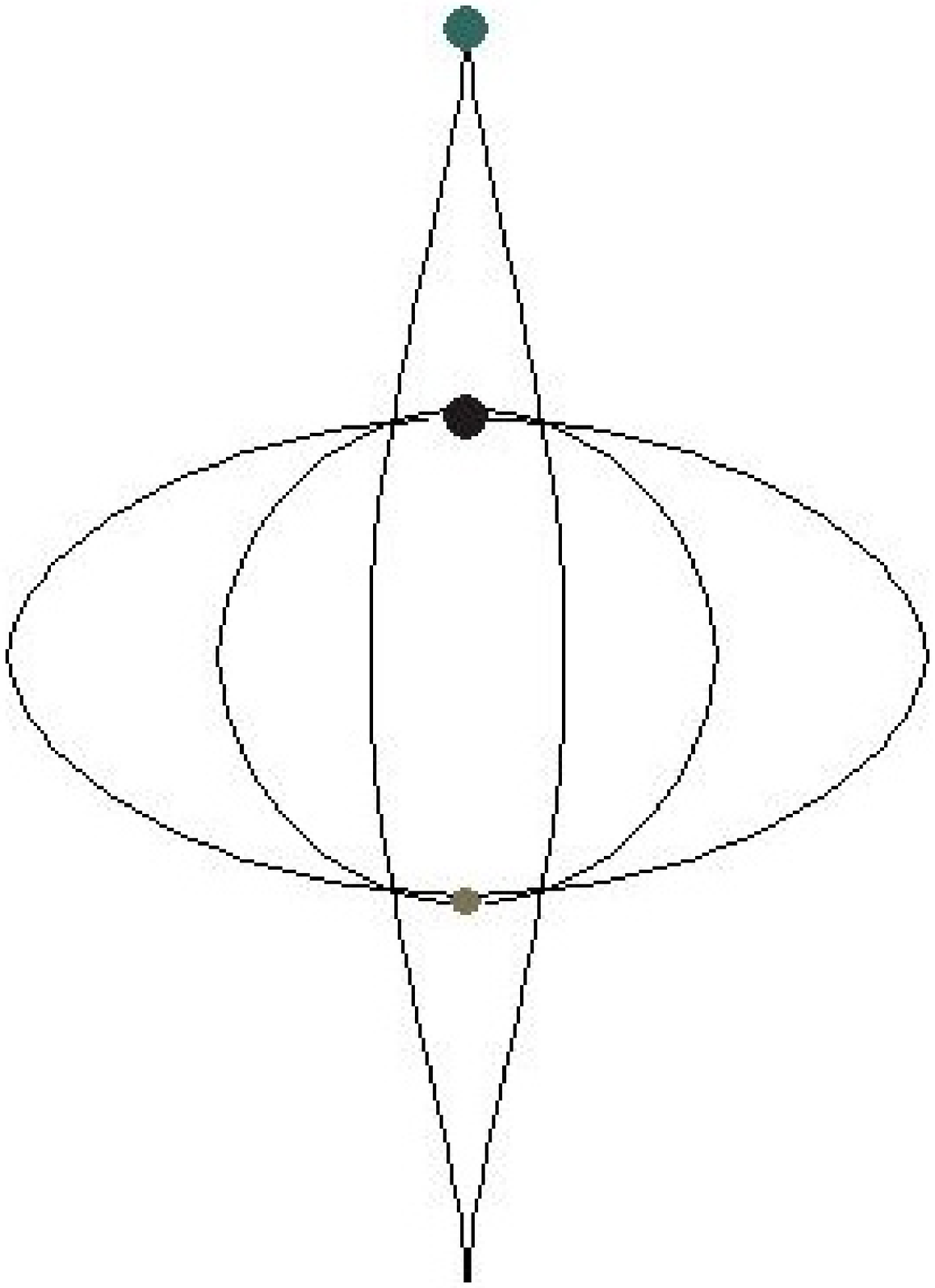} \\
Ducati3\_alluneq
\end{center}\end{minipage}
\begin{minipage}{2in} \begin{center}
\includegraphics[width=2in]{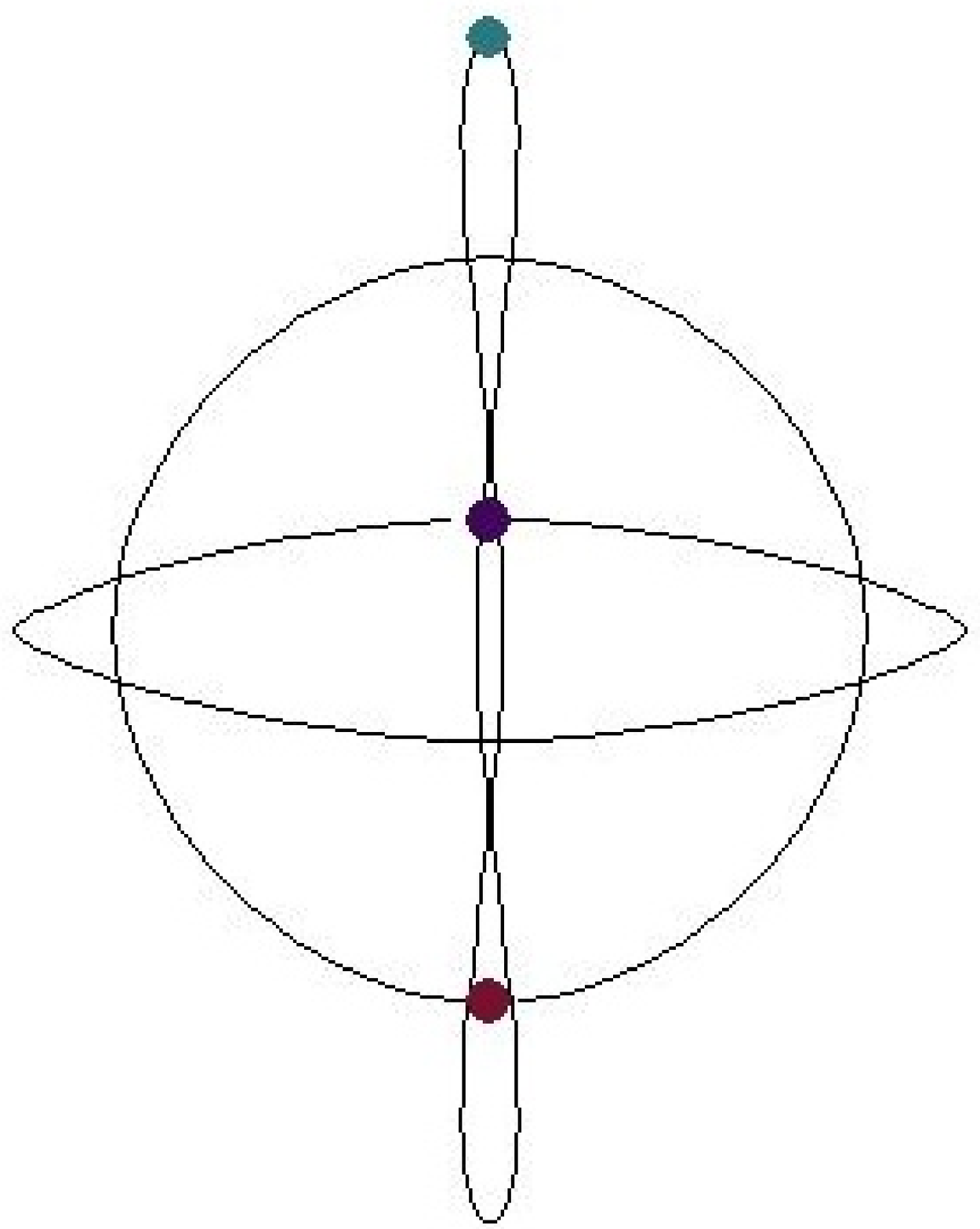} \\
Ducati3\_alluneq2
\end{center}\end{minipage}
\end{center}
\caption{Periodic Orbits---Ducati's with unequal masses.}
\label{fig:fig3}
\end{figure}

\begin{figure}
\begin{center} 

\begin{minipage}{2in} \begin{center}
\includegraphics[width=2in]{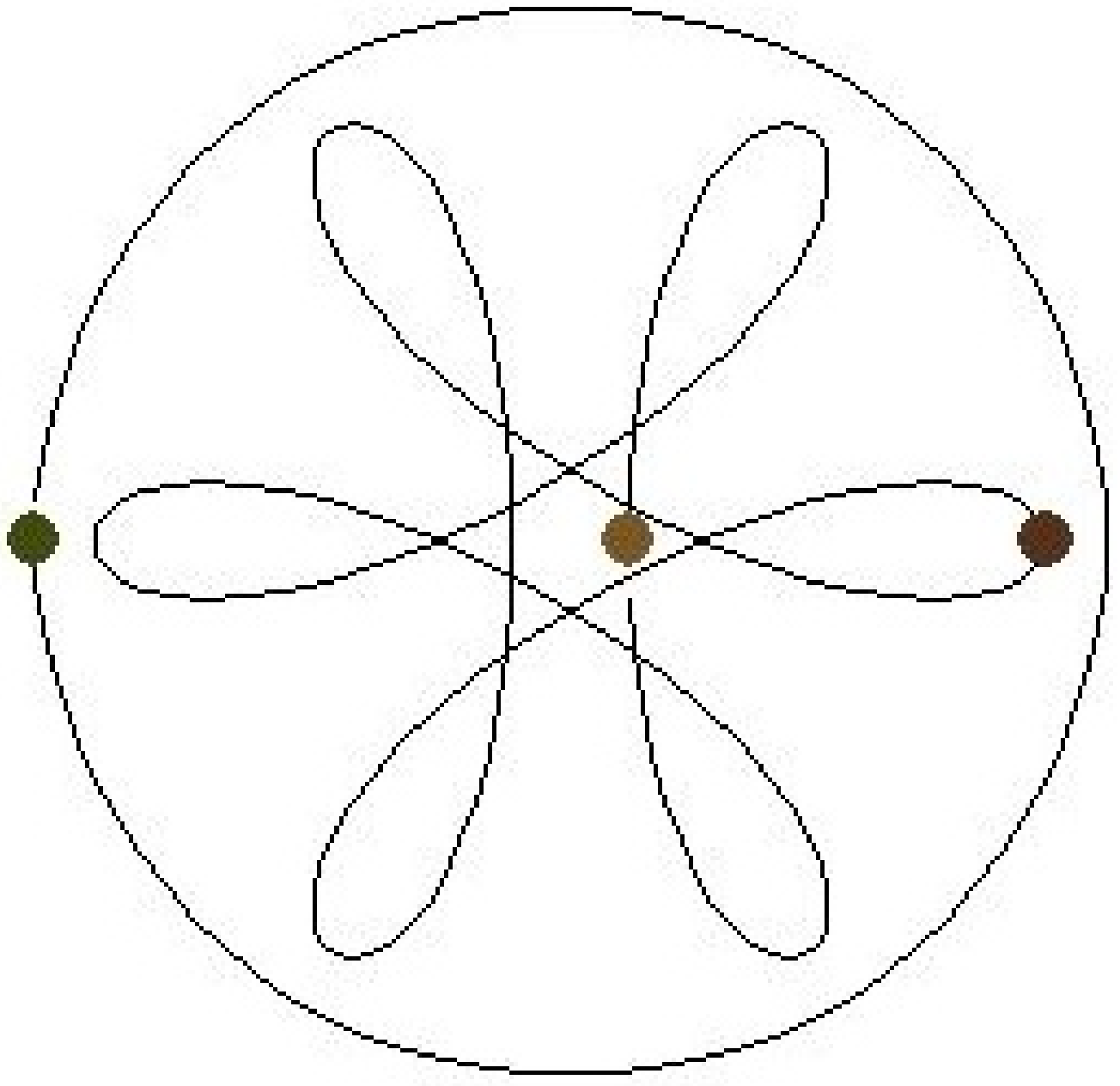} \\
Hill3\_2
\end{center}\end{minipage}
\begin{minipage}{2in} \begin{center}
\includegraphics[width=2in]{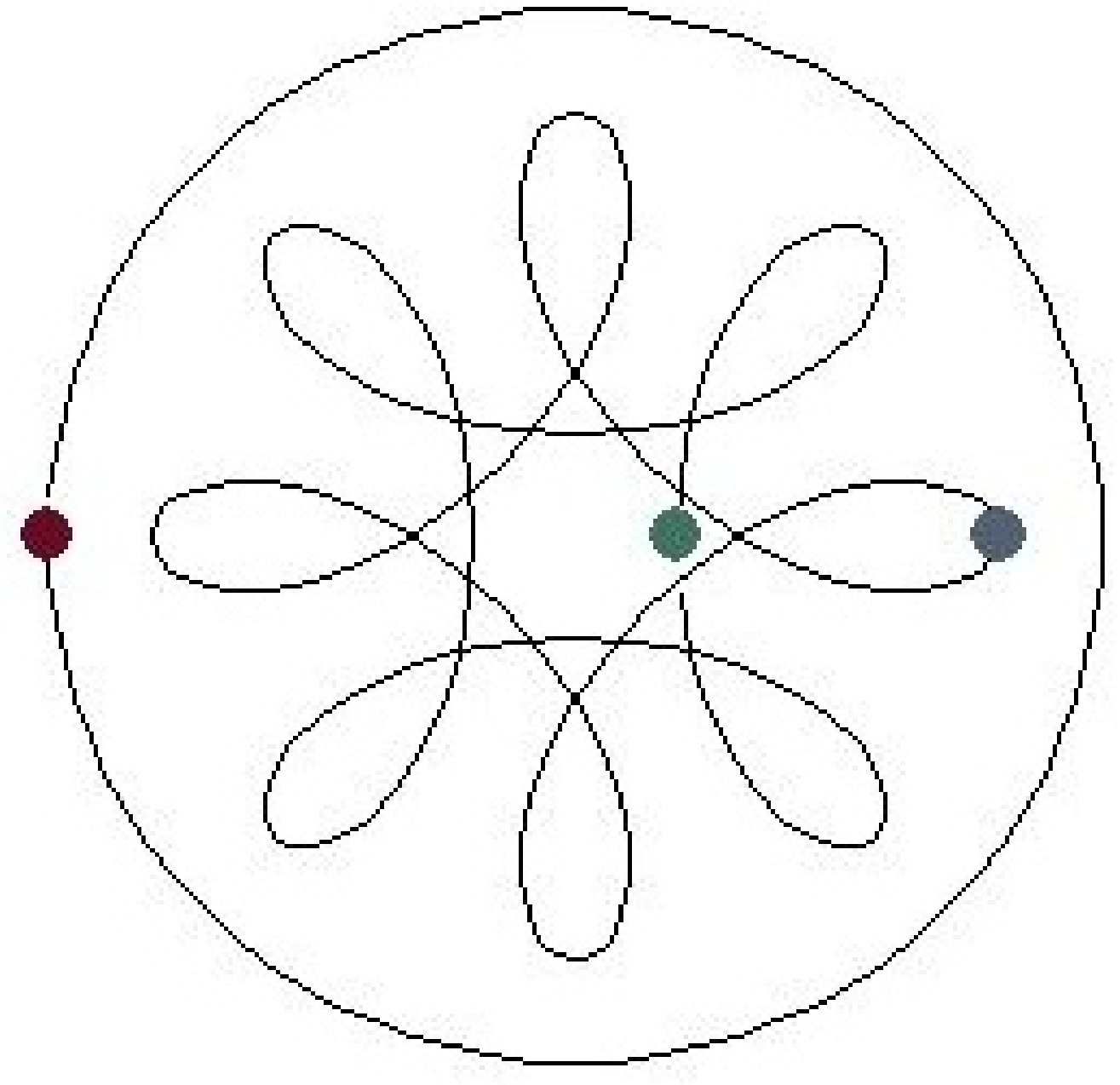} \\
Hill3\_3
\end{center}\end{minipage}
\begin{minipage}{2in} \begin{center}
\includegraphics[width=2in]{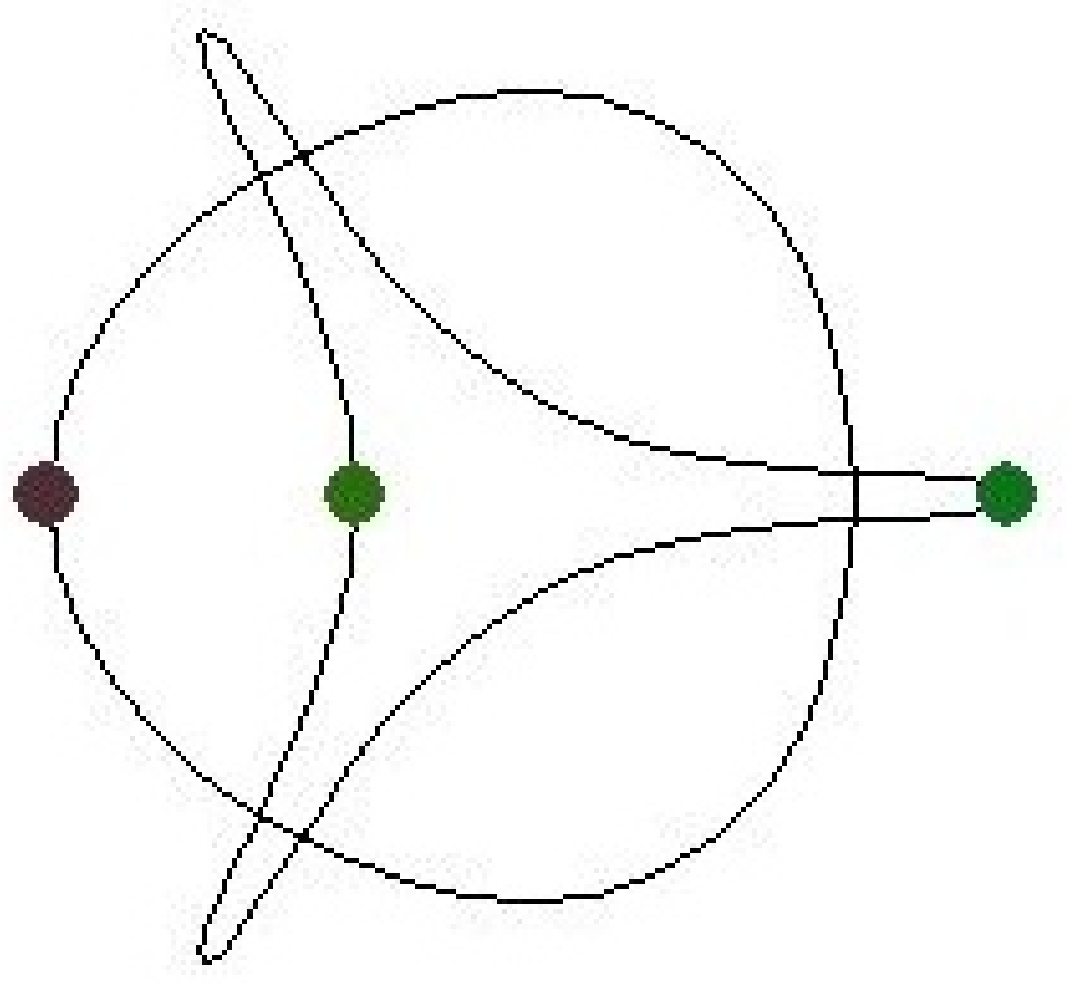} \\
Hill3\_0.5
\end{center}\end{minipage}
\end{center}
\caption{Periodic Orbits---Hill-type with equal masses.}
\label{fig:fig4}
\end{figure}

\clearpage

\end{document}